\documentclass[%
 reprint,
unsortedaddress,
footinbib,
bibnotes,
 amsmath,amssymb,
 aps,
]{revtex4-2}

\usepackage{graphicx}
\usepackage{dcolumn}
\usepackage{bm}
\usepackage{acro} 
\usepackage{xcolor}
\usepackage{physics}
\usepackage{amsmath}
\usepackage{tikz}
\usepackage{mathdots}
\usepackage{yhmath}
\usepackage{cancel}
\usepackage{color}
\usepackage{siunitx}
\usepackage{array}
\usepackage{multirow}
\usepackage{amssymb}
\usepackage{gensymb}
\usepackage{tabularx}
\usepackage{extarrows}
\usepackage{booktabs}
\usetikzlibrary{fadings}
\usetikzlibrary{patterns}
\usetikzlibrary{shadows.blur}
\usetikzlibrary{shapes}
\usepackage{mathtools, nccmath}
\usepackage{aas_macros}

\DeclareMathAlphabet\mathbfcal{OMS}{cmsy}{b}{n}
\DeclarePairedDelimiter{\nint}\lfloor\rceil
\DeclareAcronym{sa}{short = SA, long = singly-averaged}
\DeclareAcronym{da}{short = DA, long = doubly-averaged}
\DeclareAcronym{mmr}{short = MMR, long = mean-motion resonance}
\DeclareAcronym{vzlk}{short = vZLK, long = von Zeipel-Lidov-Kozai}
\DeclareAcronym{eba}{short = EBA, long = effective-binaries approximation}
\DeclareAcronym{eom}{ short = EOM, long = Equations of Motion, short-plural = s, }
\DeclareAcronym{smbh}{ short = SMBH, long = Supermassive Black Hole, short-plural = s, }
\DeclareAcronym{gw}{ short = GW, long = Gravitational Wave, short-plural =, }
\DeclareAcronym{lisa}{ short = LISA, long = Laser Interferometric  Space  Antenna, short-plural =, }
\DeclareAcronym{pn}{ short = PN, long = Post-Newtonian, short-plural =, }
\DeclareAcronym{ligo}{ short = LIGO, long = Laser Interferometer Gravitational-Wave Observatory, short-plural =, }
\DeclareAcronym{msp}{ short = MSP, long = millisecond pulsar, short-plural = s, }
\DeclareAcronym{emrb}{ short = EMRB, long = extreme-mass-ratio binary, short-plural = s, long-plural-form = extreme-mass-ratio binaries, }
\DeclareAcronym{emri}{ short = EMRI, long = extreme-mass-ratio-inspiral, short-plural = s, }
\DeclareAcronym{mpd}{ short = MPD, long = Mathisson-Papapetrou-Dixon, short-plural =, }
\DeclareAcronym{imri}{ short = IMRI, long = intermediate-mass-ratio-inspiral, short-plural = s, }
\DeclareAcronym{imrb}{ short = IMRB, long = intermediate-mass-ratio binary, short-plural = s, long-plural-form = intermediate-mass-ratio binaries, }
\DeclareAcronym{ska}{ short = SKA, long = Square Kilometer Array, }
\DeclareAcronym{adm}{ short = ADM, long = Arnowitt-Deser-Misner, }
\DeclareAcronym{bh}{short = BH, long=black hole, short-plural = s, long-plural = s}
\DeclareAcronym{tpq}{short = TPQ, long=test particle quadrupole-level approximation}

\newcommand{\solarmass}{\,{\rm M}_\odot}

\allowdisplaybreaks

\begin{document}

\preprint{APS/123-QED}

\title{Complete Hamiltonian Framework of Relativistic Hierarchical Triple Systems: \\
Capabilities and Limitations of Secular Perturbation Theory}

\author{Kaye Jiale Li}
 \affiliation{Mullard Space Science Laboratory, University College London, Holmbury St Mary, Surrey RH5 6NT, UK.}
 \affiliation{Department of Physics, Chinese University of Hong Kong, Shatin, NT, Hong Kong SAR, China.}

\author{Kinwah Wu} \author{Ziri Younsi}
 \affiliation{Mullard Space Science Laboratory, University College London, Holmbury St Mary, Surrey RH5 6NT, UK.}

\author{Tjonnie G.F. Li}
\affiliation{Laboratory for Semiconductor Physics, Department of Physics and Astronomy, KU Leuven, B-3001 Leuven, Belgium}
\affiliation{STADIUS Center for Dynamical Systems, Signal Processing and Data Analytics, Department of Electrical Engineering (ESAT), KU Leuven, B-3001 Leuven, Belgium}
\affiliation{Leuven Gravity Institute, KU Leuven, Celestijnenlaan 200D box 2415, 3001 Leuven, Belgium}

\date{\today}

\begin{abstract}
Relativistic secular perturbation theory
  has ignited significant interest
  in uncovering intricate cross-term effects,
  especially the interplay between 1PN and quadrupole terms.
While most existing studies rely on the Lagrangian planetary perturbation method
  for computing cross terms, a comprehensive Hamiltonian framework for the field has been missing.
In this work, we introduce a framework
  based on von Zeipel transformation,
  utilizing two sequential canonical transformations to
  systematically compute cross terms to arbitrary orders.
Our results reveal secular cross terms up to quadrupole-squared order,
  showcasing remarkable consistency with both
  the Lagrangian method \citep{Will2021} and
  the effective-field-theory approach \citep{Kuntz2023}.
We present leading-order periodic cross terms
  arising from the interactions between 1PN and quadrupole,
  and present estimates of higher-order cross terms.
It is demonstrated that this method not only accurately predicts the
  long-term evolution of hierarchical systems
  but also captures fast oscillations observed in N-body simulations.
We identify and validate resonances caused by quadrupole-squared effects,
  highlighting both consistencies and discrepancies when compared to N-body simulations.
These discrepancies underscore the importance of mean-motion resonances,
  a factor overlooked in current secular perturbation frameworks.
Finally, we provide a comprehensive review of the subtleties
  and limitations inherent to secular perturbation theory,
  paving the way for future research and advancements in this field.

\end{abstract}

\maketitle

\section{\label{sec:introduction}Introduction}

The dynamics of three-body systems have long intrigued researchers
  across chaos theory,
  celestial mechanics \citep[e.g.,][]{Benitez2008,Li2014,Munoz2016,Marcus2020},
  binary evolution \citep[e.g.,][]{Thompson2011,Katz2012,Haim2018},
  and gravitational wave (GW) physics \citep{Hoffman2007,Galaviz2011,Deme2020},
  with particular focus on the eccentricity excitation and oscillation in inclination
  caused by the third body.
In GW physics, third-body-induced mergers are believed to
  constitute a fraction of black hole (BH) mergers observable via
  GWs \citep{Blaes2002,Miller2002,Wen2003,Antonini2014,Hoang2018,Takatsy2019},
  where binary BH mergers
  are accelerated by eccentricity oscillations known
  as \ac{vzlk} oscillations \citep{vonZeipel1910,Brouwer1959,Lidov1962,Kozai1962,Ito2020}, a phenomenon well-explained by celestial secular perturbation theory.
Note that
  the recent discovery \citep{Peissker2024arXiv}
  of a binary system near
  Sagittarius A* suggests that stellar-mass binary systems,
  regardless of their age,
  are abundant in the vicinity of
  supermassive black holes.
The inferred age of this binary system
  also indicates the subtlety of \ac{vzlk}
  oscillations in action in such hierarchical triple systems.

Among such systems, binary-extreme-mass-ratio-inspirals (b-EMRIs)
  comprising a compact binary inspiraling
  around a supermassive BH
  are of special interest due to their dual-frequency GW emissions and
  unique signatures during stellar-mass binary mergers \citep[see, e.g.,][]{Addison2015,Chen2018a,Chen2018b,Randall2018,Gupta2020}.
The complex and intriguing dynamics of such b-EMRI systems
  have been demonstrated by \citep{Remmen2013}.
The relatively stable configuration of these hierarchical triple systems results
  in long evolution timescales, making N-body simulations computationally expensive and
  impractical for extensive parameter space exploration.
Thus, secular perturbation theory becomes essential,
  especially when augmented with relativistic corrections to model the evolution of these binaries.

Considerable effort has been dedicated to extending \ac{vzlk} theory
  into the relativistic domain \citep[see, e.g.,][]{Biscani2013,Naoz2013b,Iorio2014,Will2014a,Liu2019,Lim2020,Kuntz2021,Kuntz2023,Hamilton2024}.
The leading-order relativistic effect is generally known to
  suppress eccentricity oscillations when its time scale is shorter than
  the \ac{vzlk} oscillation \cite{Holman1997,Lin1998,Ford2000,Fabrycky2007,Dong2014,Liu2015,Will2017}.
Additional relativistic terms were derived by \citet{Naoz2013b},
  who also provided a comprehensive review of hierarchical triple systems \citep{Naoz2016}.
Extending the theory to higher orders introduces complex interactions and cross terms,
  such as the quadrupole-squared term reported by \citep{Cuk2004,Luo2016}
  in the test-particle limit and by \citet{Will2021}
  using the Lagrangian planetary perturbation approach.
\citet{Kuntz2023} also reported this term
  using the effective-field-theory approach.
The Hamiltonian approach to this cross term was first calculated by
  Brown in a series of works \citep{Brown1936A,Brown1936B,Brown1936C}
  assuming a circular outer binary.
This quadrupole-squared term was reported by
\citep{Marchal1990,Soderhjelm1975,Krymolowski1999,Breiter2015,Lei2018,Tremaine2023}
  from similar Hamiltonian approach,
  using either von Zeipel transformation or
  Lie-Hori–Deprit transformation \citep[see, e.g.,][for a review]{Shevchenko2017}.

For historical reasons the focus of study has centred on
  the test-particle regime.
Hence, the treatment of ascending nodes in many studies
  has been simplified and is not applicable to the general case.
\citet{Lei2018} presented a relatively complete framework
  for computing the quadrupole-squared cross term
  by considering the fast oscillations of angular momentum and projected angular momentum.
However, they omitted the contributions from the first canonical transformation,
  which the same author later addressed \citep{Lei2019},
  incorporating \ac{mmr} in the low-eccentricity context
  for a test particle.

More recently, \citet{Tremaine2023} cross-validated the quadrupole-squared
  Hamiltonian reported by different studies, identifying diverse definitions
  of fictitious time as the primary reason for differing results.
They provided a comprehensive review, demonstrating consistency
  across the literature and showing agreement with \citep{Soderhjelm1975,Cuk2004,Luo2016,Will2021}.
The cross term arising from the interplay of 1PN and quadrupole terms was
  reported by \citep{Will2014a,Lim2020,Kuntz2023}, though their results differ.
A closely related work by \citet{Iorio2014} derived
  the relativistic contribution in the
  test-particle limit by
  considering the coupling of the perturber's gravito-electric
  and gravito-magnetic fields with the motion of the inner binary.
These differences might stem from their use of different equations of motion,
  varying definitions of cross terms, choices of centre-of-mass,
  the distinctions between contact and osculating elements,
  and the specifics of their averaging processes.

These inconsistencies highlight the importance of clarifying
  subtle differences in assumptions,
  indicating the need for an independent, systematic approach
  to reliably calculate cross-term contributions for cross-validation.
These challenges have motivated this work, aiming to unify secular perturbation theory
  from a Hamiltonian perspective, enabling systematic calculation of
  higher-order interaction terms and
  facilitating cross-validation
  between secular perturbation theory and N-body simulations.

In this work, we present a theoretical framework based on
  von Zeipel transformation \citep{vonZeipel1910,Brouwer1959}.
We perform double-averaging (DA) of the Hamiltonian
  via two sequential canonical near-identity transformations.
Employing factors $\delta \equiv M_0/a_0$ and $\alpha \equiv a_0/a_3$
  (see definitions in Sec.~\ref{sec:canonicalTransformation})
  for two-dimensional Taylor expansion,
  we show that cross terms naturally arise in this process,
  enabling systematic calculations to arbitrary order.
We outline the procedure for computing first-order cross terms
  (i.e., two-term cross terms)
  and second-order cross terms (i.e., three-term cross terms),
  highlighting the intricate issue of angular momentum conservation
  that arises in the second-order cross terms.

In Sec.~\ref{sec:canonicalTransformation}, we perform a multipole expansion
  on the \ac{adm} Hamiltonian, assuming a hierarchical configuration.
We rewrite the Hamiltonian using Delaunay elements and then
  eliminate the mean anomalies via two sequential canonical transformations to
  obtain the \ac{da} Hamiltonian.
The secular terms are derived up to $(0,9/2)$ (i.e., $\sim \delta^0 \alpha^{9/2}$) order
  for Newtonian corrections and up to $(1,5/2)$ order
  for relativistic corrections.
Secular terms appear at $(0,3)$ (quadrupole term),
  $(0,4)$ (octupole term), $(0,9/2)$ (quadrupole-squared term) orders
  for Newtonian interactions, and at $(1,0)$ (1PN term of inner binary),
  $(1,1/2)$ (no secular effect),
  $(1,1)$ (no secular effect other than altering orbital period),
  $(1,2)$ (1PN term of outer binary; no secular effect from the interaction term),
  $(1,5/2)$ (spin-orbit coupling/de Sitter precession)
  orders for relativistic interactions.
Their formulae are reported in Appendix~\ref{ap:formula}.
Additionally, we identify two periodic cross terms at
  $(1,3/2)$ and $(1,5/2)$ orders,
  which indirectly influence the secular evolution
  by altering the correspondence between the original Hamiltonian system
  and the \ac{da} Hamiltonian system.
The secular cross term at $(0,9/2)$ order
  shows remarkable consistency with Will's work using the Lagrangian perturbation method \citep{Will2021}.
Consequently, we are also able to confirm consistency
  with Kuntz's findings from the effective-field-theory approach \citep{Kuntz2023}
  as well as with other studies in the literature \citep[see][for details]{Tremaine2023}.

In Sec.~\ref{sec:recover}, we demonstrate that fast oscillations commonly
  observed in N-body simulations,
  but absent in secular perturbation theory,
  can be recovered using the generating functions of these canonical transformations.
Our case studies in Sec.~\ref{sec:results} identify several resonances
  due to the quadrupole-squared cross term
  and confirm its physical effects.

We highlight both the alignments and mismatches between N-body simulations
  and secular perturbation theory,
  emphasising the capabilities and inherent limitations of the secular perturbation approach.
We discuss the differences between contact orbital elements
  used in Hamiltonian formulations and
  osculating orbital elements in Lagrangian formulations in Sec.~\ref{sec:CommentsLagrangian}.
We address the subtleties of eliminating ascending nodes
  and the issue of invariant reference planes in Sec.~\ref{sec:Elimination}.
Additionally, we comment on these issues in the context of gravitational radiation,
  warning against naively combining secular perturbation theory
  with radiation backreaction, highlighting potential artefacts in GW waveforms.
Higher-order terms beyond this work are estimated in Sec.~\ref{sec:HigherOrderTerms}.
Finally, limitations of secular perturbation theory are discussed in
  Sec.~\ref{sec:Limitation},
  focusing on the mean motion-related resonances that
  are absent in the current framework.

\section{\label{sec:canonicalTransformation} Canonical transformations}

\subsection{Multipole expansion of ADM Hamiltonian}

The Hamiltonian formula of a three-body system in the \ac{adm} gauge is given by \cite{Schafer1987}
\begin{align}
& \mathbb{K}_{{\rm N}}= \frac{1}{2}\sum_{a}\frac{\left|{\bf p}_{a}\right|^{2}}{m_{a}}- \frac{G}{2}\sum_{a}\sum_{b \neq a}\frac{m_{a}m_{b}}{r_{a, b}}\, , \nonumber \\
& \mathbb{K}_{1{\rm PN}}=  - \frac{1}{8}\sum_{a}m_{a}\left( \frac{\left|{\bf p}_{a}\right|^{2}}{m_{a}^{2}}\right)^{2}
- \frac{G}{4}\sum_{a}\sum_{b \neq a}\frac{1}{r_{a, b}}\nonumber  \\
& \quad
  \times \bigg[ 6 \frac{m_{b}}{m_{a}}\left|{\bf p}_{a}\right|^{2}
 - 7{\bf p}_{a}\cdot{\bf p}_{b}- \left({\bf n}_{a, b}\cdot{\bf p}_{a}\right) \left({\bf n}_{a, b}\cdot{\bf p}_{b}\right) \bigg] \nonumber \\
& \quad + \frac{G^2}{2}\sum_{a}\sum_{b \neq a}\sum_{c \neq a}\frac{m_{a}m_{b}m_{c}}{r_{a, b}r_{a, c}}\, ,
\label{eq:ADM1PN}
\end{align}
where ${\bf r}_{a,b} = {\bf r}_{a} - {\bf r}_{b}$, and ${\bf n}_{a, b} = {\bf r}_{a, b}/|{\bf r}_{a, b}|$.
Hereafter, we set $G=c=1$.
In light of the Poincar\'e symmetry,
  three conserved quantities exist at the 1PN level \cite{Landau1975,Damour2000},
   and they are
\begin{align}
&    {\bf p}_{\rm tot}  = \sum_{a} {\bf p}_{a} \ ,
  \nonumber \\
&    {\bf J}_{\rm tot}  = \sum_{a} {\bf r}_{a} \times {\bf p}_{a} \ , \nonumber  \\
&    {\bf G}_{\rm tot}  = \sum_{a} \left( m_{a} + \frac{\left| {\bf p}_{ a } \right|^{2} }{2 m_{a}}
    - \sum_{b \neq a} \frac{m_{a} m_{b}}{2 r_{a,b}} \right) {\bf r}_{a} \, ,
\label{eq:pJG}
\end{align}
  corresponding to the total momentum, the total angular momentum, and the centre-of-mass of the triple system respectively.
To account for the system consisting of an inner binary
  and an outer binary,
  we set ${\bf p}_{\rm tot} = 0$ and $\sum_{a} m_{a} {\bf r}_{a} = 0$
  (at the cost of a drifting center-of-mass at the 1PN level)
  to eliminate three coordinates and three momenta.
This facilitates the transformation of the coordinates and momenta into
  new coordinates $\mathbfcal{R}_0$, $\mathbfcal{R}_3$
  and new momenta $\mathbfcal{P}_0$, $\mathbfcal{P}_3$ via the followings:
\begin{align}
&{\bf r_{1,2}}  = \mathbfcal{R}_0 \ , \nonumber \\
&{\bf r_{1,3}}  = \frac{m_2}{m_1 + m_2} \mathbfcal{R}_0 + \mathbfcal{R}_3 \ , \nonumber  \\
&{\bf r_{2,3}}  = - \frac{m_1}{m_1 + m_2} \mathbfcal{R}_0 + \mathbfcal{R}_3 \ , \nonumber \\
&{\bf p_3}  = - \mathbfcal{P}_3 \ , \nonumber \\
&{\bf p_1}  = \mathbfcal{P}_0 +  \frac{m_1}{m_1 + m_2} \mathbfcal{P}_3 \ , \nonumber \\
&{\bf p_2}  = -\mathbfcal{P}_0 + \frac{m_2}{m_1 + m_2} \mathbfcal{P}_3 \ .
\end{align}
Here the subscript $0$
  designates the inner binary,
  composed of $m_1$ and $m_2$, and
  the subscript $3$ the outer binary.
We denote $a_0$ as the semi-major axis of the inner binary
  and $a_3$ as that of the outer binary.
The reduced mass and total mass are defined as
  $\mu_0 = m_1 m_2/M_0$, $M_0 = m_1 + m_2$, $\mu_3 = m_3 M_0/ M_3$ and
  $M_3 = m_3 + M_0$.
Further, we define $\Delta m = m_2 - m_1 \ge 0$.
The specifics here are chosen for b-EMRI systems,
  but generalizing to hierarchical triple systems of arbitrary mass ratio is straightforward.
For a hierarchical triple system,
  the two factors $\alpha = a_0/a_3$ and $\delta = M_0/a_0$
  are much smaller than unity, allowing the Hamiltonian to be expanded
  with respect to these factors:
\begin{align}
\mathcal{K}  = \mathbb{K}_{{\rm N}} + \mathbb{K}_{1{\rm PN}}
 = \sum_{n,m} \mathcal{K}_{{\rm int},n,\frac{m}{2}} + \sum_{n,m} \mathcal{K}_{{\rm b},n,\frac{m}{2}} \ ,
\end{align}
with
\begin{align}
\mathcal{K}_{{\rm int},n,\frac{m}{2}} & \sim \mathcal{K}_{{\rm b},n,\frac{m}{2}}
    \sim  \mathcal{K}_{{\rm b},0,0} \delta^n \alpha^{\frac{m}{2}}
    \sim a_0^{\frac{m}{2}-n-1} a_3^{-\frac{m}{2}} \ . \nonumber
\end{align}
We use $\mathcal{K}_{{\rm int},n,m/2}$ to represent the interaction terms and $\mathcal{K}_{{\rm b},n,m/2}$ for the binary terms,
  referring to these collectively as the Hamiltonian of $(n,m/2)$ order.
$\mathcal{K}_{{\rm int},0,3}$ and $\mathcal{K}_{{\rm int},0,4}$ are the well-known
  Newtonian quadrupole and octupole interaction terms.
$\mathcal{K}_{{\rm b},0,0}$ and $\mathcal{K}_{{\rm b},1,0}$ represent the
  Newtonian and 1PN Hamiltonian terms of the inner binary.
 $\mathcal{K}_{{\rm b},0,1}$ and $\mathcal{K}_{{\rm b},1,2}$ represent the corresponding terms for the outer binary.
This new Hamiltonian has its angular momentum,
 $\mathbfcal{J} \equiv \sum_{a} \mathbfcal{R}_{a} \times \mathbfcal{P}_{a}$, conserved.

The Hamiltonian terms are given by
\begin{widetext}
\begin{align}
&\mathcal{K}_{{\rm b},0,0}  = \frac{{\mathcal{P}_0}^2}{2 \mu_0}-\frac{\mu_0 M_0}{\mathcal{R}_0} \ , \, \text{and $(0 \to 3)$ for the outer binary } \mathcal{K}_{{\rm b},0,1}\  , \nonumber \\
&\mathcal{K}_{{\rm b},1,0}  = \frac{3 {\mathcal{P}_0}^4}{8 {\mu_0}^2 M_0}-\frac{3 M_0 {\mathcal{P}_0}^2}{2 \mu_0 \mathcal{R}_0}+\frac{\mu_0 {M_0}^2}{2 {\mathcal{R}_0}^2}-\frac{{\mathcal{P}_0}^4}{8 {\mu_0}^3}-\frac{{\mathcal{P}_0}^2}{2 \mathcal{R}_0}-\frac{{\mathcal{P}_0}^2 \cos^2 \Phi_0}{2 \mathcal{R}_0}  \ , \, \text{and $(0 \to 3)$ for outer binary } \mathcal{K}_{{\rm b},1,2}\ , \nonumber \\
&\mathcal{K}_{{\rm int},0,3}  = -\frac{m_1 m_2  m_3 {\mathcal{R}_0}^2 }{4 M_0 {\mathcal{R}_3}^3} (3 \cos 2 \psi +1)
\ , \nonumber \\
&\mathcal{K}_{{\rm int},0,4}  = \frac{\Delta m m_1 m_2 m_3 {\mathcal{R}_0}^3 }{8 {M_0}^2 {\mathcal{R}_3}^4} (3 \cos \psi +5 \cos 3 \psi )
\ , \nonumber  \\
&\mathcal{K}_{{\rm int},1,\frac{1}{2}}  = \frac{\Delta m \mathcal{P}_0 \mathcal{P}_3 }{2 {\mu_0}^2 {M_0}^2 \mathcal{R}_0}
\left({\mu_0}^2 M_0 \left(\cos \theta +\cos \beta  \cos \Phi_0 \right)-{\mathcal{P}_0}^2 \mathcal{R}_0 \cos \theta \right)
\ , \nonumber \\
&\mathcal{K}_{{\rm int},1,1}  = \frac{\left(2 m_3 {M_0}^2 \left(4 {\mu_0}^2 M_0-3 {\mathcal{P}_0}^2 \mathcal{R}_0\right)+{\mathcal{P}_3}^2 \mathcal{R}_3 \left({\mu_0}^2 M_0 (\cos 2 \beta +3)-{\mathcal{P}_0}^2 \mathcal{R}_0 (\cos 2 \theta +2)\right)\right)}{4 \mu_0 {M_0}^2 \mathcal{R}_0 \mathcal{R}_3} \ ,  \nonumber \\
&\mathcal{K}_{{\rm int},1,2}  = \frac{\Delta m m_3  \left(3 {\mathcal{P}_0}^2 \mathcal{R}_0-2 {\mu_0}^2 M_0\right)}{2 \mu_0 M_0 {\mathcal{R}_3}^2} \cos \psi  \ , \nonumber \\
&\mathcal{K}_{{\rm int},1,\frac{5}{2}}   = -\frac{\mathcal{P}_0 \mathcal{P}_3 \mathcal{R}_0}{2 {M_0}^2 {\mathcal{R}_3}^2} \bigg\{  {M_0}^2 \big[ -7 \cos \theta   \cos \psi  +\cos \beta   \cos \epsilon  +\cos \varPhi_3  (\cos \varPhi_0 -3 \cos \psi   \cos \epsilon  ) \big] -6 \mu_3 M_3 \cos \theta   \cos \psi  \bigg\}
\ ,
\end{align}
\end{widetext}
where $\beta$ is the angle between $\mathbfcal{R}_0$ and $\mathbfcal{P}_3$,
  $\theta$ is the angle between $\mathbfcal{P}_0$ and $\mathbfcal{P}_3$,
  $\psi$ is the angle between $\mathbfcal{R}_0$ and $\mathbfcal{R}_3$.
  $\epsilon$ is the angle between $\mathbfcal{P}_0$ and $\mathbfcal{R}_3$,
  $\Phi_0$ is the angle between $\mathbfcal{P}_0$ and $\mathbfcal{R}_0$, and
  $\Phi_3$ is the angle between $\mathbfcal{P}_3$ and $\mathbfcal{R}_3$.
We have included all the terms up to $(0,4)$ order for the Newtonian Hamiltonian
  and up to $(1,5/2)$ order for the relativistic Hamiltonian,
  omitting all trivial terms.
Simulations performed using this Hamiltonian are referred to as \ac{eba}.
Note that $\mathcal{K}_{{\rm int},1,5/2}$ is of the same order as spin-orbit coupling.

\subsection{Delaunay elements}

\begin{figure}
\vspace*{-1cm} \centering
\includegraphics[width=1\columnwidth]{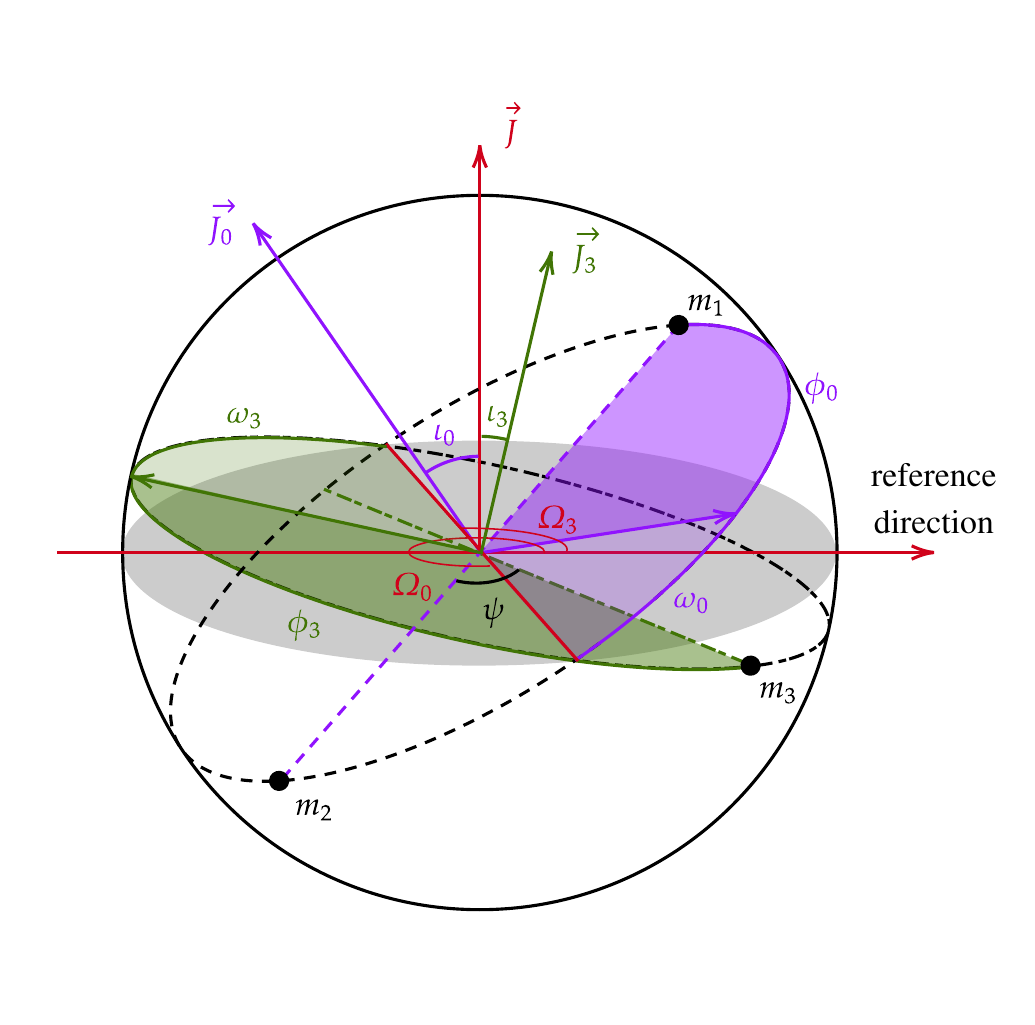}
\vspace*{-1cm}
  \caption{
  The diagram illustrates the geometry of the triple system.
  The reference plane is represented by the shaded circular area in grey.
  The centers of mass for both the inner and outer binaries are positioned at the origin
    of the reference coordinate system, and the relative positions
    of the three masses are projected onto the surface of a unit sphere.
  $\Omega_{0} (\Omega_3)$ represents the longitude of the ascending node.
  $\omega_{0}$ ($\omega_3$) is the argument of periapsis of the inner (outer) binary,
   depicted by the lighter teal (magenta) sector within the orbital plane,
    while the true anomaly $\phi_{0}$ ($\phi_3$) is depicted by the darker teal (magenta) sector.}
  \label{fig:geometry}
\end{figure}

For a hierarchical triple system,
  it is usually advantageous to convert the position coordinates and momenta into Delaunay elements \citep{Valtonen2006}.
We denote the new coordinates as ${\boldsymbol Q} \equiv ({\boldsymbol Q}_0,{\boldsymbol Q}_3)$
  and their conjugate momenta as ${\boldsymbol P} \equiv ({\boldsymbol P}_0,{\boldsymbol P}_3)$.
Specifically, ${\boldsymbol Q}_0 = (l_0, \omega_0, \Omega_0)$ and
  ${\boldsymbol P}_0 = (L_0, J_0, H_0)$, with similar definitions for the outer binary.
Here $l_0$ is the mean anomaly,
  $\omega_0$ is the argument of periapsis,
  $\Omega_0$ is longitude of the ascending node,
  $J_0$ denotes the angular momentum, and $H_0$ is the projected angular momentum
  of the inner binary in the direction of total angular momentum
  ${\boldsymbol J} = {\boldsymbol J}_0 + {\boldsymbol J}_3$,
  as shown in Fig.~\ref{fig:geometry}.
We define $\iota = \iota_0 + \iota_3$ as the total inclination angle,
  which is equivalent to the relative inclination angle between two orbital planes
  if and only if $\delta \Omega \equiv \Omega_0 - \Omega_3 = \pi$.
The old coordinates and their conjugate momenta are related to
  the new coordinates and their conjugate momenta via
  the following:
  (similarly for the outer binary)
\begin{widetext}
\begin{align}
& {\boldsymbol R}_0 =
{\rm Rotation}~[\Omega_0,\iota_0,\omega_0] \times
\big( R_0 \cos \phi_0, R_0 \sin \phi_0, 0 \big) \ , \nonumber \\
& {\boldsymbol P}_0 = {\rm Rotation}~[\Omega_0,\iota_0,\omega_0] \times
\bigg( P_{r,0} \cos \phi_0 - \frac{J_0}{R_0} \sin \phi_0,
P_{r,0} \sin \phi_0 + \frac{J_0}{R_0} \cos \phi_0, 0 \bigg) \ ,
 \nonumber \\
& R_0 = \frac{a_0 (1 - {e_0}^2)}{1 + e_0 \cos \phi_0} \ ,
 \nonumber \\
& P_0 = \pm \frac{\mu_0}{R_0} \sqrt{\frac{M_0}{a_0}
   \bigg[ {a_0}^2 {e_0}^2-\left(a_0-R_0\right)^2 \bigg]}\ ,
   \nonumber \\
& P_{r,0} = \frac{{\mu_0}^2 M_0}{J_0} e_0 \sin \phi_0 \ ,
\nonumber \\
& \phi_0 = l_0 + 2\sum _{s=1}^{\infty }{\frac {1}{s}}\left[J_{s}(s e_0)+\sum _{p=1}^{\infty }\chi^{p}{\big(}J_{s-p}(se_0)+J_{s+p}(s e_0){\big)}\right]\sin(sl_0)  \  \quad
\left(
\text{with} \,  \chi \equiv \frac{1-{\sqrt {1-{e_0}^{2}}}}{e_0}
\right) \ , \nonumber \\
& L_0 = \mu_0 \sqrt{M_0 a_0} \ , \nonumber \\
& J_0 = L_0  \sqrt{1 - {e_0}^2} \ , \nonumber \\
& H_0 = J_0 \cos \iota_0 \ .
\end{align}
Here $J_s$ is Bessel function of the
  first kind \citep{Brouwer1961}
   and the rotation matrix
   ${\rm Rotation}~[\Omega_0,\iota_0,\omega_0]$
   is given by
\begin{equation}
\begin{aligned}
& {\rm Rotation}~[\Omega_0,\iota_0,\omega_0] = \\
& \hspace*{1cm}
\left(
\begin{array}{ccc}
 \cos \omega_0 \cos \Omega_0-\cos \iota_0 \sin \omega_0 \sin \Omega_0 & - \sin \omega_0 \cos \Omega_0 -\cos \iota_0 \cos \omega_0 \sin \Omega_0 & \sin \iota_0 \sin \Omega_0 \\
 \cos \iota_0 \sin \omega_0 \cos \Omega_0+\cos \omega_0 \sin \Omega_0 & \cos \iota_0 \cos \omega_0 \cos \Omega_0-\sin \omega_0 \sin \Omega_0 & - \sin \iota_0 \cos \Omega_0 \\
 \sin \iota_0 \sin \omega_0 & \sin \iota_0 \cos \omega_0 & \cos \iota_0 \\
\end{array}
\right) \,,
\end{aligned}
\end{equation}
\end{widetext}
This transformation preserves the canonical structure of
  the Hamiltonian.
The details of which can be found in, for example, \cite{Valtonen2006}.

Using Delaunay elements,
  the new Hamiltonian can be expressed as
\begin{align}
K (\boldsymbol{Q}, \boldsymbol{P}) & =
K (l_0, \omega_0, \Omega_0, l_3, \omega_3, \Omega_3, L_0, J_0, H_0, L_3, J_3, H_3) \nonumber \\
& = \sum_{n,m} K_{n,\frac{m}{2}} \ ,
\end{align}
where
$K_{n,\frac{m}{2}}  \sim K_{0,0} \delta^n \alpha^{\frac{m}{2}}
    \sim {a_0}^{\frac{m}{2}-n-1} {a_3}^{-\frac{m}{2}} \ . $
Here, we have grouped binary and interaction terms together
  in the expression for uniformity.
The first two terms originating from the Newtonian binary terms are independent of mean anomalies:
\begin{align}
&K_{0,0}\ (L_0) = -\frac{{\mu_0}^3 {M_0}^2}{2 {L_0}^2} \ ,
\nonumber \\
&K_{0,1}\ (L_3) = -\frac{{\mu_3}^3 {M_3}^2}{2 {L_3}^2} \ .
\end{align}
The evolution equations are then
\begin{align}
&\frac{{\rm d} l_0}{{\rm d} t} = - \frac{\partial K}{\partial L_0}\ , \,\; \frac{{\rm d} L_0}{{\rm d} t} = \frac{\partial K}{\partial l_0} \ , \,\;
{\rm and \,} (0 \to 3) \ , \nonumber \\
&\frac{{\rm d} \omega_0}{{\rm d} t} = - \frac{\partial K}{\partial J_0} \ , \,\; \frac{{\rm d} J_0}{{\rm d} t} = \frac{\partial K}{\partial \omega_0} \ , \,\; {\rm and \,} (0 \to 3) \ ,
  \nonumber \\
&\frac{{\rm d} \Omega_0}{{\rm d} t} = - \frac{\partial K}{\partial H_0} \ , \,\; \
\frac{{\rm d} H_0}{{\rm d} t} = \frac{\partial K}{\partial \Omega_0} \ , \,\; {\rm and \,} (0 \to 3) \ .
\end{align}
In the absence of interaction terms and relativistic binary terms,
  the mean anomalies $l_0$ and $l_3$ simply grow linearly with time,
  while all other elements remain constants of motion.

\subsection{First non-trivial canonical transformation}
\label{sec:FirstCanonical}

As our focus lies on the long-term evolution of the triple system,
  our objective is to eliminate the fast oscillating components
  (i.e., $l_0$ and $l_3$) of the Hamiltonian.
This procedure effectively averages the Hamiltonian
  over the two orbital periods, leaving only the slowly evolving secular terms.
We follow the work of Brouwer \cite{Brouwer1959} to
  address the secular evolution
  arising from the multipole mass interactions and relativistic effects.
The first canonical transformation aims at eliminating $l_0$ and leaving the outer orbit intact.
This can be achieved by a near-identity canonical transformation,
  maintaining the canonical structure while ensuring the canonical coordinates and momenta
  closely resemble their original counterparts.
We denote the new canonical coordinates and momenta
  $(\boldsymbol{Q^{\ast}},\boldsymbol{P^{\ast}})$
  with $\boldsymbol{Q^{\ast}} \equiv (\boldsymbol{Q^{\ast}}_0, \boldsymbol{Q^{\ast}}_3)$,
  $\boldsymbol{P^{\ast}} \equiv (\boldsymbol{P^{\ast}}_0, \boldsymbol{P^{\ast}}_3)$
  where $\boldsymbol{Q^{\ast}}_0 \equiv (l_0^{\ast}, \omega_0^{\ast}, \Omega_0^{\ast})$,
        $\boldsymbol{P^{\ast}}_0 \equiv (L_0^{\ast}, J_0^{\ast}, H_0^{\ast})$, and similarly for the outer binary.
Given the utility of definitions such as semi-major axis,
  eccentricity, and inclination during calculations,
   we refer to the semi-major axis defined based on $L_0^{\ast}$ as $a_0^{\ast}$,
  the eccentricity defined based on $L_0^{\ast}$ and $J_0^{\ast}$ as $e_0^{\ast}$,
  and similarly for $a_3^{\ast}$, $e_3^{\ast}$, $\iota_0^{\ast}$ and $\iota_3^{\ast}$.
We denote the total inclination angle as $\iota^{\ast} = \iota_0^{\ast} + \iota_3^{\ast}$ as before.

Suppose that the near-identity transformation from $(\boldsymbol{Q},\boldsymbol{P})$
  to $(\boldsymbol{Q^{\ast}},\boldsymbol{P^{\ast}})$
  can be achieved with the aid of a generating function
  $S(\boldsymbol{Q},\boldsymbol{P^{\ast}})$, if
\begin{align}
&L_0  = \frac{\partial S}{\partial l_0}, \,\; J_0 = \frac{\partial S}{\partial \omega_0}, \,\; H_0 = \frac{\partial S}{\partial \Omega_0} , \,\; \text{and $(0 \to 3)$} \ ,    \nonumber \\
&l_0^{\ast}  = \frac{\partial S}{\partial L_0^{\ast}}, \,\; \omega_0^{\ast} = \frac{\partial S}{\partial J_0^{\ast}}, \,\; \Omega_0^{\ast} = \frac{\partial S}{\partial H_0^{\ast}} , \,\; \text{and $(0 \to 3)$} \ .
\label{eq:nearIdentityQPRelation1}
\end{align}
The new Hamiltonian remains canonical, with
\begin{align}
\label{eq:KKstarRelation1}
K (l_0,\omega_0, \Omega_0,\boldsymbol{Q}_3,\boldsymbol{P})  = K^{\ast} (-,\omega_0^{\ast}, \Omega_0^{\ast}, \boldsymbol{Q^{\ast}}_3,\boldsymbol{P^{\ast}})\ ,
\end{align}
where a dashed line is added to emphasize that $K^{\ast}$ becomes independent of $l_0^{\ast}$.
We will henceforth refer to $K^{\ast}$ as {\it \ac{sa}} Hamiltonian,
  and  $(\boldsymbol{Q^{\ast}},\boldsymbol{P^{\ast}})$ as SA orbital elements.
As suggested by \cite{Brouwer1959},
  the generating function can be expressed as a Taylor series:
\begin{align}
\label{eq:SDef}
S(\boldsymbol{Q},\boldsymbol{P^{\ast}})  = \boldsymbol{Q} \cdot \boldsymbol{P^{\ast}}
 +  \sum_{n,m} S_{n,\frac{m}{2}}
   (\boldsymbol{Q},\boldsymbol{P^{\ast}}) \ ,
\end{align}
 where
\begin{align}
S_{n,\frac{m}{2}}
   (\boldsymbol{Q},\boldsymbol{P^{\ast}})  \sim \frac{1}{K'_{0,0}(L_0)}  \delta^n \alpha^{\frac{m}{2}} \sim {a_0}^{\frac{m}{2}-n+\frac{1}{2}} {a_3}^{-\frac{m}{2}}\ .    \nonumber
\end{align}
Given that $\delta \ll 1 $ and $\alpha \ll 1$, the summation over $S_{n,m/2}$
  is a small perturbation.
This construction of generating function ensures that $({\boldsymbol Q}, {\boldsymbol P})$
  only differ from $({\boldsymbol Q^{\ast}}, {\boldsymbol P^{\ast}})$ by a small perturbation, hence the term ``near-identity''.

Replacing the $\boldsymbol{P}$ and $\boldsymbol{Q^{\ast}}$ in Eq.~\eqref{eq:KKstarRelation1}
  using the relation Eq.~\eqref{eq:nearIdentityQPRelation1},
  we obtain
\begin{align}
& K\left(l_0, \omega_0, \Omega_0, \boldsymbol{Q}_3,
\frac{\partial S}{\partial \boldsymbol{Q}}\right)  \nonumber \\
 & \hspace*{1cm}
 = K^{\ast}\left(-,  \frac{\partial S}{\partial J_0^{\ast}} , \frac{\partial S}{\partial H_0^{\ast}} , \frac{\partial S}{\partial \boldsymbol{P^{\ast}}_3} ,\boldsymbol{P^{\ast}}\right) \ .
\end{align}
Both sides can be expanded with respect to the two small variables $\delta$ and $\alpha$.
At each  $(n,m/2)$ order, we have
\begin{widetext}
\begin{align}
K^{\ast}_{(i),n,\frac{m}{2}} = \frac{\partial K_{0,0} }{\partial L_0^{\ast}} \frac{\partial S_{(i),n,\frac{m}{2}} }{\partial l_0}
  + C_{(i),n,\frac{m}{2}}
  \hspace*{0.5cm}
  \left({\rm for} \, i = {\rm b},0,1
   \right)
   \ ,
\label{eq:FirstCanonicalK}
\end{align}
with
\begin{align}
& C_{(0),n,\frac{m}{2}} = K_{n,\frac{m}{2}}  \ , \nonumber \\
& C_{(1),n,\frac{m}{2}}
  = \sum_{k,l,|k|+|l|\neq 0} \bigg(
  \frac{\partial C_{(0),k,\frac{l}{2}}}{\partial \boldsymbol{P^{\ast}}_0}
          \frac{\partial S_{(0),n-k,\frac{m-l}{2}}}{\partial \boldsymbol{Q}_0}
 - \frac{\partial K^{\ast}_{(0),k,\frac{l}{2}}}{\partial \boldsymbol{Q}_0}
          \frac{\partial S_{(0),n-k,\frac{m-l}{2}}}{\partial \boldsymbol{P^{\ast}}_0}
+ \frac{\partial C_{(0),k,\frac{l}{2}}}{\partial \boldsymbol{P^{\ast}}_3}
          \frac{\partial S_{(0),n-k,\frac{m-l-1}{2}}}{\partial \boldsymbol{Q}_3} \nonumber \\
& \hspace*{1cm}
 - \frac{\partial K^{\ast}_{(0),k,\frac{l}{2}}}{\partial \boldsymbol{Q}_3}
          \frac{\partial S_{(0),n-k,\frac{m-l-1}{2}}}{\partial \boldsymbol{P^{\ast}}_3}
+ \frac{1}{2} \frac{\partial^2 K_{0,0}}{\partial L^{\ast}_0{}^2}
          \frac{\partial S_{(0),k,\frac{l}{2}}}{\partial l_0} \frac{\partial S_{(0),n-k,\frac{m-l}{2}}}{\partial l_0}
+ \frac{1}{2} \frac{\partial^2 K_{0,1}}{\partial L^{\ast}_3{}^2}
          \frac{\partial S_{(0),k,\frac{l}{2}}}{\partial l_3} \frac{\partial S_{(0),n-k-i,\frac{m-l}{2}-2}}{\partial l_3}
 \bigg)
 \, . \nonumber
\end{align}
\end{widetext}
The new total Hamiltonian becomes
\begin{align}
K^{\ast} &  =
    \sum_{n,m} K^{\ast}_{n,\frac{m}{2}} \nonumber \\
   &  = \sum_{n,m} \bigg( K^{\ast}_{({\rm b}),n,\frac{m}{2}} + K^{\ast}_{(0),n,\frac{m}{2}} + K^{\ast}_{(1),n,\frac{m}{2}} \bigg) \ .
\end{align}
Here, $K^{\ast}_{({\rm b})}$ represents the binary terms.
$K^{\ast}_{(0)}$ represents the SA term that follows directly from $K$,
  and will be referred to as {\it direct term}.
$K^{\ast}_{(1)}$ represents the {\it cross term}
  that arises from the interplay of two direct terms.

The generating function $S_{(i),n,m/2}$ at each order are selected to
  satisfy the equality, capturing the periodic component, leaving $K^{\ast}_{(i)}$ free from $l_0^{\ast}$.
It follows
\begin{align}
&K^{\ast}_{(i),n,\frac{m}{2}}  =
  \big\langle C_{(i),n,\frac{m}{2}} \big\rangle_{l_0} \equiv
  \frac{1}{2 \pi} \int_{0}^{2\pi} {\rm d}l_0\, C_{(i),n,\frac{m}{2}} \ ,  \nonumber \\
&S_{(0),n,\frac{m}{2}}  =
  \frac{1}{K'_{0,0}(L_0^{\ast})} \bigg\{ \int {\rm d}l_0\,
    \big\{ C_{(0),n,\frac{m}{2}} \big\}_{l_0}  \bigg\}_{l_0} \ ,
\end{align}
where we have used the notation $\{ \cdots \}_{l_0} $ to represent the zero-mean component
  with respect to $l_0$:
\begin{align}
& \big\{ C_{(i),n,\frac{m}{2}} \big\}_{l_0}  \equiv
    C_{(i),n,\frac{m}{2}} - \left< C_{(i),n,\frac{m}{2}} \right>_{l_0} \ .
\end{align}
We omit $S_{(1),n,m/2}$ here as they would only be necessary for cross terms of higher order.
Thus, the secular components of the original Hamiltonian
  are captured by $K^{\ast}_{(i),n,m/2}$, while the periodic components are absorbed into $S_{(i),n,m/2}$.
It is worth noticing that we have
\begin{align}
\left\langle
S_{(0),n,\frac{m}{2}}
\right\rangle_{l_0}
& = \left\langle
 \frac{\partial S_{(0),n,\frac{m}{2}} }{\partial \boldsymbol{P^{\ast}}}
\right\rangle_{l_0} \nonumber \\
& = \left\langle
\frac{\partial S_{(0),n,\frac{m}{2}} }{\partial \boldsymbol{Q}} \right\rangle_{l_0}
 = 0 \ .
\end{align}
This condition ensures that there
  is no secular difference between the old coordinates and the new coordinates at this order,
  although this argument entails some subtleties, as discussed in Sec.~\ref{sec:Limitation}.


Truncating the series above $(0,9/2)$ order for Newtonian contribution
  and $(1,5/2)$ order for relativistic corrections, the direct terms appear at
  $(0,3)$, $(0,4)$, $(1,0)$, $(1,1/2)$, $(1,1)$, $(1,2)$, $(1,5/2)$ orders.
Among these direct terms, $K^{\ast}_{(0),1,1/2}$ vanishes,
  implying that this term only has periodic contributions to the system.
The trivial leading orders are simply
\begin{equation}
\begin{aligned}
K^{\ast}_{({\rm b}),0,0} & = K_{0,0}(L_0^{\ast}) \,\, \;\! {\rm and}  \,\;\!
  K^{\ast}_{({\rm b}),0,1} = K_{0,1}(L_3^{\ast}) \ .
\end{aligned}
\end{equation}
Equation~\eqref{eq:FirstCanonicalK} suggests that
\begin{equation}
\begin{aligned}
C_{(i),n,\frac{m}{2}} & \sim K^{\ast}_{(i),n,\frac{m}{2}}
    \sim {a_0}^{\frac{m}{2}-n-1} {a_3}^{-\frac{m}{2}} \ , \\
S_{(i),n,\frac{m}{2}} & 
    \sim {a_0}^{\frac{m}{2}-n+\frac{1}{2}} {a_3}^{-\frac{m}{2}} \,. \\
\end{aligned}
\end{equation}
Hence the leading order Newtonian term appears at $(0,6)$ order
  due to the self-interaction of the quadrupole term.
The leading order relativistic cross term arises from the
  interaction of the 1PN inner binary term
  with the quadrupole term, appearing at $(1,3)$ order.
Both of these cross terms are ignored in this study.

\subsection{Second non-trivial canonical transformation}
To eliminate $l_3^{\ast}$ from $(\boldsymbol{Q^{\ast}},\boldsymbol{P^{\ast}})$,
  we perform the second canonical transformation to yield
   a new set of coordinates and conjugate momenta $(\boldsymbol{Q^{\ast\ast}},\boldsymbol{P^{\ast\ast}})$.
We refer to  $(\boldsymbol{Q^{\ast\ast}},\boldsymbol{P^{\ast\ast}})$ as the {\it \acl{da}} orbital elements,
  and the new Hamiltonian as the DA Hamiltonian.
The generating function
  of this near-identity canonical transformation reads
\begin{align}
\label{eq:SastDef}
S^{\ast} (\boldsymbol{Q^{\ast}},\boldsymbol{P^{\ast\ast}})  = \boldsymbol{Q^{\ast}} \cdot \boldsymbol{P^{\ast\ast}}
 +  \sum_{n,m} S^{\ast}_{n,\frac{m}{2}} (\boldsymbol{Q^{\ast}},\boldsymbol{P^{\ast\ast}}) \ ,
\end{align}
where
\begin{align}
S^{\ast}_{n,\frac{m}{2}} (\boldsymbol{Q^{\ast}},\boldsymbol{P^{\ast\ast}})
  \sim \frac{1}{K'_{0,1}(L_3)} \delta^n \alpha^{\frac{m}{2}}
  \sim {a_0}^{\frac{m}{2}-n-1} {a_3}^{\frac{3-m}{2}} \ .
 \nonumber
\end{align}
Similar to the previous section,
  the old and new coordinates and momenta are related by
\begin{align}
\label{eq:nearIdentityQPRelation2}
&L^{\ast}_0  = \frac{\partial S^{\ast}}{\partial l^{\ast}_0},
    \,\; J^{\ast}_0 = \frac{\partial S^{\ast}}{\partial \omega^{\ast}_0},
    \,\; H^{\ast}_0 = \frac{\partial S^{\ast}}{\partial \Omega^{\ast}_0} , \,\; \text{and $(0 \to 3)$,}
     \nonumber \\
&l_0^{\ast\ast}  = \frac{\partial S^{\ast}}{\partial L_0^{\ast\ast}},
    \,\; \omega_0^{\ast\ast} = \frac{\partial S^{\ast}}{\partial J_0^{\ast\ast}},
    \,\; \Omega_0^{\ast\ast} = \frac{\partial S^{\ast}}{\partial H_0^{\ast\ast}} , \,\; \text{and $(0 \to 3)$}\, .
\end{align}
We again refer to the new semi-major axes, eccentricities, and inclination angles
  defined with respect to $(\boldsymbol{Q^{\ast\ast}},\boldsymbol{P^{\ast\ast}})$
  as $e_{0}^{\ast\ast}$, $a_{0}^{\ast\ast}$ and $\iota_{0}^{\ast\ast}$, etc.
The new Hamiltonian is related to the old Hamiltonian via
\begin{align}
\label{eq:KKstarRelation}
&K^{\ast} (-,\omega_0^{\ast}, \Omega_0^{\ast}, l_3^{\ast},\omega_3^{\ast}, \Omega_3^{\ast},\boldsymbol{P^{\ast}})      =  \nonumber \\
  & \hspace*{1cm} K^{\ast\ast} (-,\omega_0^{\ast\ast}, \Omega_0^{\ast\ast}, -,\omega_3^{\ast\ast}, \Omega_3^{\ast\ast},\boldsymbol{P^{\ast\ast}})\ .
\end{align}

\begin{widetext}
Rewriting this relation in terms of $(\boldsymbol{Q^{\ast}},\boldsymbol{P^{\ast\ast}})$ yields
\begin{align}
K^{\ast\ast}_{(i),n,\frac{m}{2}} =
  \frac{\partial K^{\ast}_{0,1} }{\partial L^{\ast\ast}_3} \frac{\partial S^{\ast}_{(i),n,\frac{m}{2}} }{\partial l^{\ast}_3}
  + C^{\ast}_{(i),n,\frac{m}{2}}
\label{eq:SecondCanonicalK}
\end{align}
(for $i = {\rm b},0,1,2$), where
\begin{align}
&C^{\ast}_{(0),n,\frac{m}{2}}   = K^{\ast}_{n,\frac{m}{2}}  \ ,
  \nonumber \\
&C^{\ast}_{(1),n,\frac{m}{2}}   = \sum_{k,l,(k,l) \neq (0,2)}
  \bigg(
  \frac{\partial C^{\ast}_{(0),k,\frac{l}{2}}}{\partial \boldsymbol{P^{\ast\ast}}_0}
          \frac{\partial S^{\ast}_{(0),n-k,\frac{m-l+3}{2}}}{\partial \boldsymbol{Q^{\ast}}_0}
- \frac{\partial K^{\ast\ast}_{(0),k,\frac{l}{2}}}{\partial \boldsymbol{Q^{\ast}}_0}
          \frac{\partial S^{\ast}_{(0),n-k,\frac{m-l+3}{2}}}{\partial \boldsymbol{P^{\ast\ast}}_0}
+ \frac{\partial C^{\ast}_{(0),k,\frac{l}{2}}}{\partial \boldsymbol{P^{\ast\ast}}_3}
          \frac{\partial S^{\ast}_{(0),n-k,\frac{m-l}{2}+1}}{\partial \boldsymbol{Q^{\ast}}_3} \nonumber \\
& \hspace*{2cm} -\frac{\partial K^{\ast\ast}_{(0),k,\frac{l}{2}}}{\partial \boldsymbol{Q^{\ast}}_3}
          \frac{\partial S^{\ast}_{(0),n-k,\frac{m-l}{2}+1}}{\partial \boldsymbol{P^{\ast\ast}}_3}
+ \frac{1}{2} \frac{\partial^2 K^{\ast}_{0,1}}{\partial L^{\ast\ast}_3{}^2}
          \frac{\partial S^{\ast}_{(0),k,l}}{\partial l^{\ast}_3}
          \frac{\partial S^{\ast}_{(0),n-k-i,\frac{m-l}{2}+1}}{\partial l^{\ast}_3}
 \bigg)  \ ,  \nonumber \\
&C^{\ast}_{(2),n,\frac{m}{2}} = \sum_{k,l,(k,l) \neq (0,2)}
  \bigg(
  \frac{\partial C^{\ast}_{(0),k,\frac{l}{2}}}{\partial \boldsymbol{P^{\ast\ast}}_0}
          \frac{\partial S^{\ast}_{(1),n-k,\frac{m-l+3}{2}}}{\partial \boldsymbol{Q^{\ast}}_0}
- \frac{\partial K^{\ast\ast}_{(0),k,\frac{l}{2}}}{\partial \boldsymbol{Q^{\ast}}_0}
          \frac{\partial S^{\ast}_{(1),n-k,\frac{m-l+3}{2}}}{\partial \boldsymbol{P^{\ast\ast}}_0}
- \frac{\partial K^{\ast\ast}_{(1),n-k,\frac{m-l+3}{2}}}{\partial \boldsymbol{Q^{\ast}}_0}
          \frac{\partial S^{\ast}_{(0),k,\frac{l}{2}}}{\partial \boldsymbol{P^{\ast\ast}}_0} \nonumber \\
& \hspace*{2cm}
+ \frac{\partial C^{\ast}_{(0),k,\frac{l}{2}}}{\partial \boldsymbol{P^{\ast\ast}}_3}
          \frac{\partial S^{\ast}_{(1),n-k,\frac{m-l}{2}+1}}{\partial \boldsymbol{Q^{\ast}}_3}
- \frac{\partial K^{\ast\ast}_{(0),k,\frac{l}{2}}}{\partial \boldsymbol{Q^{\ast}}_3}
          \frac{\partial S^{\ast}_{(1),n-k,\frac{m-l}{2}+1}}{\partial \boldsymbol{P^{\ast\ast}}_3}
- \frac{\partial K^{\ast\ast}_{(1),n-k,\frac{m-l}{2}+1}}{\partial \boldsymbol{Q^{\ast}}_3}
          \frac{\partial S^{\ast}_{(0),k,\frac{l}{2}}}{\partial \boldsymbol{P^{\ast\ast}}_3}
 \bigg) \nonumber \\
& \hspace*{2cm}
 + \sum_{k,l,i,j,(i,j) \neq (0,2)} \bigg(
\frac{1}{2} \frac{\partial^2 C^{\ast}_{(0),i,\frac{j}{2}}}{\partial \boldsymbol{P^{\ast\ast}}_0 \partial \boldsymbol{P^{\ast\ast}}_0}
    \frac{\partial S^{\ast}_{(0),k,\frac{l}{2}}}{\partial \boldsymbol{Q^{\ast}}_0} \frac{\partial S^{\ast}_{(0),n-k-i,\frac{m-l-j}{2}+3}}{\partial \boldsymbol{Q^{\ast}}_0} \nonumber \\
& \hspace*{2cm}
 + \frac{\partial^2 C^{\ast}_{(0),i,\frac{j}{2}}}{\partial \boldsymbol{P^{\ast\ast}}_0 \partial \boldsymbol{P^{\ast\ast}}_3}
    \frac{\partial S^{\ast}_{(0),k,\frac{l}{2}}}{\partial \boldsymbol{Q^{\ast}}_0} \frac{\partial S^{\ast}_{(0),n-k-i,\frac{m-l-j+5}{2}}}{\partial \boldsymbol{Q^{\ast}}_3}
+  \frac{1}{2} \frac{\partial^2 C^{\ast}_{(0),i,\frac{j}{2}}}{\partial \boldsymbol{P^{\ast\ast}}_3 \partial \boldsymbol{P^{\ast\ast}}_3}
    \frac{\partial S^{\ast}_{(0),k,\frac{l}{2}}}{\partial \boldsymbol{Q^{\ast}}_3} \frac{\partial S^{\ast}_{(0),n-k-i,\frac{m-l-j}{2}+2}}{\partial \boldsymbol{Q^{\ast}}_3} \nonumber \\
& \hspace*{2cm}
-  \frac{1}{2} \frac{\partial^2 K^{\ast\ast}_{i,\frac{j}{2}}}{\partial \boldsymbol{Q^{\ast}}_0 \partial \boldsymbol{Q^{\ast}}_0}
     \frac{\partial S^{\ast}_{(0),k,\frac{l}{2}}}{\partial \boldsymbol{P^{\ast\ast}}_0}
     \frac{\partial S^{\ast}_{(0),n-k-i,\frac{m-l-j}{2}+3}}{\partial \boldsymbol{P^{\ast\ast}}_0}
-  \frac{\partial^2 K^{\ast\ast}_{i,\frac{j}{2}}}{\partial \boldsymbol{Q^{\ast}}_0 \partial \boldsymbol{Q^{\ast}}_3}
     \frac{\partial S^{\ast}_{(0),k,\frac{l}{2}}}{\partial \boldsymbol{P^{\ast\ast}}_0}
     \frac{\partial S^{\ast}_{(0),n-k-i,\frac{m-l-j+5}{2}}}{\partial \boldsymbol{P^{\ast\ast}}_3}
\nonumber \\
& \hspace*{2cm}
-  \frac{1}{2} \frac{\partial^2 K^{\ast\ast}_{i,\frac{j}{2}}}{\partial \boldsymbol{Q^{\ast}}_3 \partial \boldsymbol{Q^{\ast}}_3}
     \frac{\partial S^{\ast}_{(0),k,\frac{l}{2}}}{\partial \boldsymbol{P^{\ast\ast}}_3}
     \frac{\partial S^{\ast}_{(0),n-k-i,\frac{m-l-j}{2}+2}}{\partial \boldsymbol{P^{\ast\ast}}_3}
+ \frac{1}{6} \frac{\partial^3 K^{\ast}_{0,1}}{\partial L^{\ast\ast}_3{}^3}
        \frac{\partial S^{\ast}_{(0),i,\frac{j}{2}}}{\partial l^{\ast}_3}
        \frac{\partial S^{\ast}_{(0),k,\frac{l}{2}}}{\partial l^{\ast}_3}
        \frac{\partial S^{\ast}_{(0),n-k-i,\frac{m-l-j}{2}+2}}{\partial l^{\ast}_3} 
  \bigg)
 \ . \nonumber
\end{align}
\end{widetext}

The new total Hamiltonian then becomes
\begin{align}
K^{\ast\ast} & =
    \sum_{n,m} K^{\ast\ast}_{n,\frac{m}{2}}  \nonumber \\
  & = \sum_{n,m} \bigg( K^{\ast\ast}_{(0),n,\frac{m}{2}} + K^{\ast\ast}_{(1),n,\frac{m}{2}} + K^{\ast\ast}_{(2),n,\frac{m}{2}} \bigg) \, .
\end{align}
Here $K^{\ast\ast}_{(0),n,m/2}$ represent a combination
  of direct terms and the cross terms that arise from
  the first canonical transformation.
Similar to before, $K^{\ast\ast}_{(1),n,m/2}$ denotes the two-term cross terms,
  and $K^{\ast\ast}_{(2),n,m/2}$ represents the three-term cross terms,
  which result from the interplay of three direct terms
  or the interaction of a direct term with a two-term cross term.
The DA Hamiltonian is related to the SA Hamiltonian via
\begin{widetext}
\begin{align}
K^{\ast\ast}_{(i),n,\frac{m}{2}} & =
  \left\langle C^{\ast}_{(i),n,\frac{m}{2}} \right\rangle_{l^{\ast}_3} \equiv
  \frac{1}{2 \pi} \int_{0}^{2\pi} {\rm d}l^{\ast}_3\, C^{\ast}_{(i),n,\frac{m}{2}} \ , \nonumber \\
S^{\ast}_{(0),n,\frac{m}{2}}   & =
  \frac{1}{K^{\ast\prime}_{0,1}(L^{\ast\ast}_3)} \left\{ \int {\rm d}l^{\ast}_3\,
    \big\{ C^{\ast}_{(0),n,\frac{m}{2}} \big\}_{l^{\ast}_3}  \right\}_{l^{\ast}_3} \ ,  \nonumber \\
S^{\ast}_{(1),n,\frac{m}{2}}   & =
  \frac{1}{K^{\ast\prime}_{0,1}(L^{\ast\ast}_3)} \left\{ \int {\rm d}l^{\ast}_3\,
    \big\{ C^{\ast}_{(1),n,\frac{m}{2}} \big\}_{l^{\ast}_3}  \right\}_{l^{\ast}_3}  +
\frac{1}{2} \sum_{k,l} \Bigg\langle
\frac{\partial S^{\ast}_{(0),k,\frac{l}{2}}}{\partial J^{\ast\ast}_0}
          \frac{\partial S^{\ast}_{(0),n-k,\frac{m-l+3}{2}}}{\partial \omega^{\ast}_0}
          + \frac{\partial S^{\ast}_{(0),k,\frac{l}{2}}}{\partial H^{\ast\ast}_0}
          \frac{\partial S^{\ast}_{(0),n-k,\frac{m-l+3}{2}}}{\partial \Omega^{\ast}_0}
 \nonumber  \\
& \hspace*{2cm}
+ \frac{\partial S^{\ast}_{(0),k,\frac{l}{2}}}{\partial J^{\ast\ast}_3}
          \frac{\partial S^{\ast}_{(0),n-k,\frac{m-l}{2}+1}}{\partial \omega^{\ast}_3}
+ \frac{\partial S^{\ast}_{(0),k,\frac{l}{2}}}{\partial H^{\ast\ast}_3}
          \frac{\partial S^{\ast}_{(0),n-k,\frac{m-l}{2}+1}}{\partial \Omega^{\ast}_3}
\Bigg\rangle_{l^{\ast}_3}  \ .
\label{eq:secondDirect}
\end{align}
\end{widetext}
It is noteworthy that our $S^{\ast}_{(1),n,m/2}$ has a non-zero mean value,
  which introduces an offset between the SA orbital elements
  and the DA orbital elements.
This seemingly unnatural choice turns out to be essential for
  preserving angular momentum conservation in its simplest form
  for the three-term cross term $K^{\ast\ast}_{(2),n,m/2}$.
A detailed discussion can be found in Appendix~\ref{ap:angularMomentum}.

The non-trivial secular direct terms appear at
  $(0,3)$, $(0,4)$, $(1,0)$, $(1,1)$, $(1,2)$, $(1,5/2)$ order.
Among the two contributions to $(1,2)$ order,
  the interaction term originating from $\mathcal{K}_{{\rm int},1,2}$ vanishes
  during this canonical transformation.
This leaves us with only the 1PN relativistic
  term of the outer binary, the sole secular term at this order.
The direct term at $(1,1)$ order has no secular effect other than altering
  the orbital periods.
These results align with that of \citep{Will2014a}.
The cross terms appear
  at $(0,9/2)$, $(1,3/2)$, and $(1,5/2)$ orders:
\begin{widetext}
\begin{align}
\label{eq:C09}
C^{\ast}_{(1),0,\frac{9}{2}} & =
\frac{\partial C^{\ast}_{(0),0,3}}{\partial \boldsymbol{P^{\ast\ast}}_0} \frac{\partial S^{\ast}_{(0),0,3}}{\partial \boldsymbol{Q^{\ast}}_0} - \frac{\partial K^{\ast\ast}_{(0),0,3}}{\partial \boldsymbol{Q^{\ast}}_0} \frac{\partial S^{\ast}_{(0),0,3}}{\partial \boldsymbol{P^{\ast\ast}}_0}
  \ , \nonumber   \\
C^{\ast}_{(1),1,\frac{3}{2}}&  =
    \frac{\partial C^{\ast}_{(0),1,0}}{\partial \boldsymbol{P^{\ast\ast}}_0}
       \frac{\partial S^{\ast}_{(0),0,3}}{\partial \boldsymbol{Q^{\ast}}_0}
  - \frac{\partial K^{\ast\ast}_{(0),1,0}}{\partial \boldsymbol{Q^{\ast}}_0}
       \frac{\partial S^{\ast}_{(0),0,3}}{\partial \boldsymbol{P^{\ast\ast}}_0}  \ ,  \nonumber \\
C^{\ast}_{(1),1,\frac{5}{2}} &  =
\frac{\partial C^{\ast}_{(0),1,0}}{\partial \boldsymbol{P^{\ast\ast}}_0} \frac{\partial S^{\ast}_{(0),0,4}}{\partial \boldsymbol{Q^{\ast}}_0}
 - \frac{\partial K^{\ast\ast}_{(0),1,0}}{\partial \boldsymbol{Q^{\ast}}_0} \frac{\partial S^{\ast}_{(0),0,4}}{\partial \boldsymbol{P^{\ast\ast}}_0}
 + \frac{\partial C^{\ast}_{(0),1,1}}{\partial \boldsymbol{P^{\ast\ast}}_0} \frac{\partial S^{\ast}_{(0),0,3}}{\partial \boldsymbol{Q^{\ast}}_0}
+ \frac{\partial C^{\ast}_{(0),0,3}}{\partial \boldsymbol{P^{\ast\ast}}_0} \frac{\partial S^{\ast}_{(0),1,1}}{\partial \boldsymbol{Q^{\ast}}_0}
  \nonumber \\
& \hspace*{2cm}
- \frac{\partial K^{\ast\ast}_{(0),1,1}}{\partial \boldsymbol{Q^{\ast}}_0} \frac{\partial S^{\ast}_{(0),0,3}}{\partial \boldsymbol{P^{\ast\ast}}_0}
- \frac{\partial K^{\ast\ast}_{(0),0,3}}{\partial \boldsymbol{Q^{\ast}}_0} \frac{\partial S^{\ast}_{(0),1,1}}{\partial \boldsymbol{P^{\ast\ast}}_0} \ ,
\end{align}
\end{widetext}
where the time average of the latter two cross terms vanishes.
The only non-zero secular cross term is  $K^{\ast\ast}_{(1),0,9/2}$.
This particular cross term is also referred to as the quadrupole-squared term,
  and our result matches exactly with that derived by \citet{Will2021}.
This alignment suggests consistency with \citep{Kuntz2023},
  who independently verified their results to be consistent with Will's findings.
Additionally, \citet{Tremaine2023}, who referred to this term
  as the Brown Hamiltonian,
  examined the consistency of the quadrupole-squared term
  across the literature.
They confirmed that the difference between
  the findings of \citet{Luo2016} and \citet{Will2021}
  is solely due to the choice of fictitious time.
The $(1,3/2)$ cross term exhibits periodic nature without secular effects,
  due to the neglect of the inner binary's relativistic precession
  over the orbital time scale of the outer binary,
  which is invalid near resonance \citep{Kuntz2022}.
This aligns with the conclusion of \citep{Kuntz2023},
  confirming that the secular term at this order, as derived by \cite{Lim2020}, should not exist.
Yet it is worth noting that in \citep{Kuntz2023},
  such a term already appears in the Lagrangian before DA,
  which is subsequently eliminated during averaging over the inner orbital period.
In contrast, our $(1,3/2)$ term arises from the interaction
  between 1PN and quadrupole effects during averaging over the outer orbital period.
This discrepancy may stem from differences in the equation of motion
  or the choice of centre-of-mass.
The complete formulae for these secular direct terms and cross terms
  are presented in Appendix~\ref{ap:formula}.
The periodic terms (i.e., the generating functions) are detailed
  in the supplementary material in the form of a {\it Mathematica} script.

\subsection{Recovering the true orbital elements}
\label{sec:recover}


The interaction and relativistic terms
  induce deviations in the orbits of both binaries from a Keplerian ellipse.
Through the canonical transformation,
  these non-Keplerian features are divided into a secular component,
  captured by $K^{\ast\ast}_{(i),n,m/2}$,
  and a periodic component, captured by $S_{(i),n,m/2}$ and $S^{\ast}_{(i),n,m/2}$.
In this process, not only are the fast oscillations of
  the Hamiltonian eliminated,
  but also those of the orbital elements are averaged out, leaving only the secular evolution.
This subtle treatment addresses
  a key disparity between the predictions of N-body simulations
  and secular perturbation theory, as highlighted by \citep[e.g.,][]{Krymolowski1999,Kinoshita2007,Luo2016,Nie2019}.
Notably, the fast eccentricity oscillations
  on orbital time scale have been shown to
  affect GW merger time scales \citep{Antognini2014}
  and tidal disruption event rates \citep{Ivanov2005}.

To facilitate the comparison between N-body simulation and that of secular perturbation theory,
   the conversion between the orbital elements and DA elements
   can be achieved using the generating function via Eq.~\eqref{eq:nearIdentityQPRelation1}
  and Eq.~\eqref{eq:nearIdentityQPRelation2},
  which we explicitly state here:
\begin{align}
\label{eq:PQrelation}
  {\boldsymbol P} & = {\boldsymbol P^{\ast\ast}} + \sum_{n,m,i}
   \left( \frac{\partial S_{(i),n,\frac{m}{2}}}{\partial {\boldsymbol Q}}
         \bigg|_{{\boldsymbol Q}, {\boldsymbol P^{\ast}}}
      +  \frac{\partial S^{\ast}_{(i),n,\frac{m}{2}}}{\partial {\boldsymbol Q^{\ast}}}
         \bigg|_{{\boldsymbol Q^{\ast}}, {\boldsymbol P^{\ast\ast}}} \right) \ , \nonumber \\
  {\boldsymbol Q} & = {\boldsymbol Q^{\ast\ast}} - \sum_{n,m,i}
    \left(
      \frac{\partial S_{(i),n,\frac{m}{2}}}{\partial {\boldsymbol P^{\ast}}}
        \bigg|_{{\boldsymbol Q}, {\boldsymbol P^{\ast}}}  +
       \frac{\partial S^{\ast}_{(i),n,\frac{m}{2}}}{\partial {\boldsymbol P^{\ast\ast}}}
        \bigg|_{{\boldsymbol Q^{\ast}}, {\boldsymbol P^{\ast\ast}}} \right) \ .
\end{align}
These equations are solved using multidimensional root-finding methods,
  specifically leveraging
  {\it gsl\_multiroot\_fsolver\_hybrids} implemented in the GNU Scientific Library.
Since both $S_{(i),n,m/2}$ and $S^{\ast}_{(i),n,m/2}$ appear as
  perturbations, it is also feasible to replace the unknown
  coordinates/momenta within these generating functions
  with known ones, and then iteratively solve for the unknown coordinates/momenta.
This iterative approach has demonstrated high efficiency
  and computational cost-effectiveness in practice.

\begin{figure*}[htp]
\centering
\includegraphics[width=1.8\columnwidth]{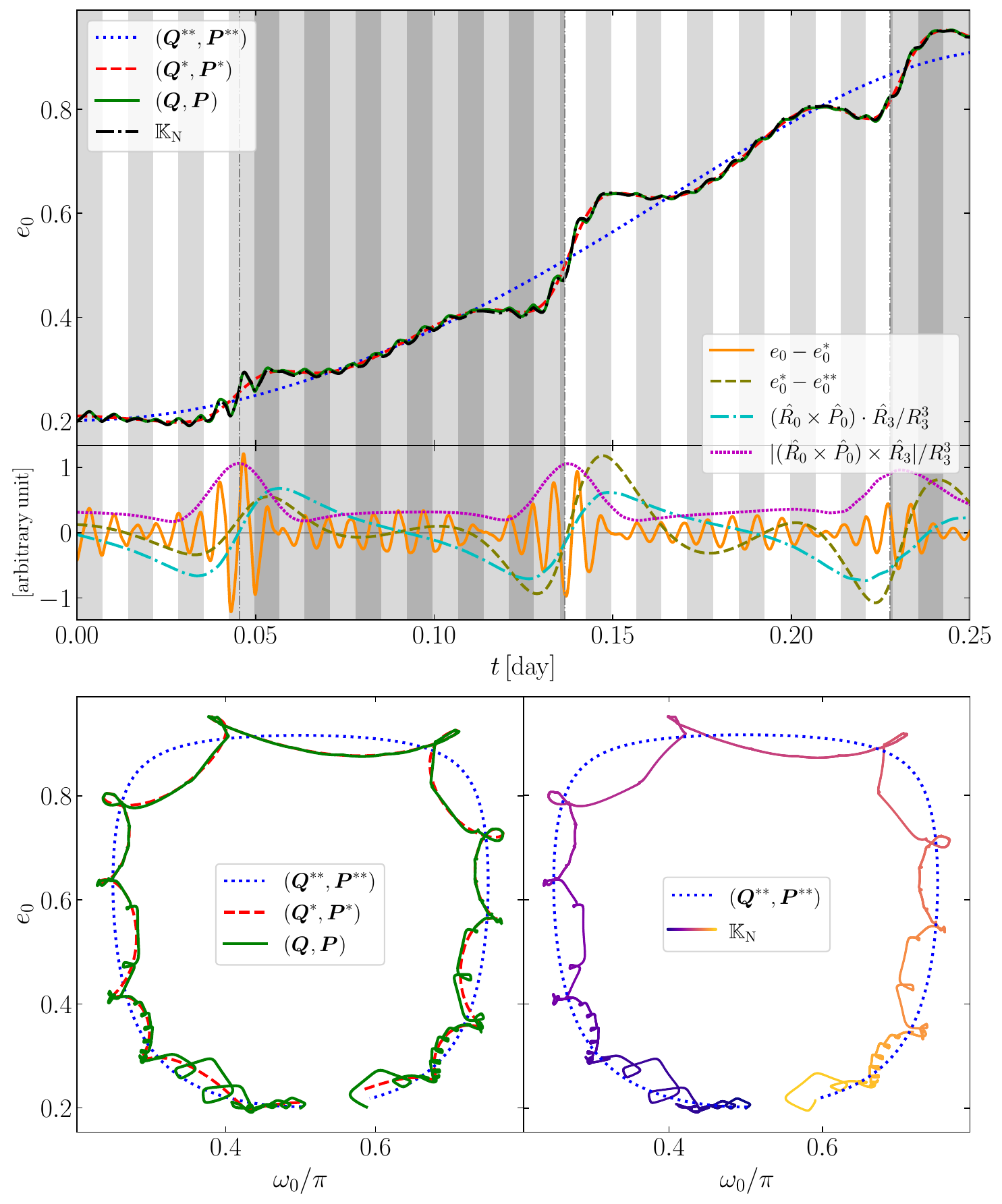}
\caption{
\label{fig:OscillatingComponents}
The triple system is composed of $m_1 = 10 \solarmass$,
   $m_2 = 10 \solarmass$
  with a $m_3 = 10^4 \solarmass$ perturber.
The initial orbital parameters are $a_0 = 10^4 (m_1 + m_2)$,
  $\iota = 110^{\circ}$,
  $a_3 = 43.298 a_0$, $\omega_0 = \pi/2$, $\omega_3 = 0.1$,
  $e_0 = 0.2$, $e_3 = 0.6$,
  $l_0 = 0$ and $l_3 = \pi$.
The orbital period of the inner binary is
   $P_0 \approx 0.0072  \,{\rm day}$
  and the orbital period of the outer binary is
   $P_3 \approx 0.091 \,{\rm day}$.
The quadrupole period is $0.59 \,{\rm day}$.
In the top two panels,
  the shaded areas represent the time intervals
  corresponding to the orbital period of the inner binary (in grey)
  and outer binary (in darker grey).
The shading is applied every other orbital period,
  representing the time interval between periapsis and the next periapsis.
The first panel shows the evolution of $e_0^{\ast\ast}$ (blue dotted),
  $e_0^{\ast}$ (red dashed), $e_0$ (green solid)
  using the DA Hamiltonian $K^{\ast\ast} = K^{\ast\ast}_{(0),0,3} + K^{\ast\ast}_{(1),0,9/2}$.
The black dashed-dotted line represents the true evolution of the
  triple system under the Hamiltonian $\mathbb{K}_{\rm N}$.
The middle panel shows the difference between
  $e_0^{} - e_0^{\ast}$ (orange solid line),
  $e_0^{\ast} - e_0^{\ast\ast}$ (olive dashed line),
  and the component of tidal force perpendicular (cyan dashed-dotted line) to
  and parallel (magenta dashed line) to
  the orbital plane of the inner binary.
All these values in the middle panel are multiplied
  by a different value to scale them into roughly the same
  number, in order for a clear visualization.
A horizontal line of $0$ is added for reference.
The bottom panels show the phase space structure
  of the \ac{da} Hamiltonian system
  (with legend the same as the upper panel),
  and the exact triple system under $\mathbb{K}_{\rm N}$
    (colourful line in the bottom right panel).
The blue dashed line is duplicated on the bottom right panel for reference.
As demonstrated in this figure,
  after correcting for the offset in initial parameters,
  the evolution of the triple system predicted by
  the DA Hamiltonian closely aligns with the results from N-body simulations.}
\end{figure*}

Figure~\ref{fig:OscillatingComponents} illustrates
  the oscillation and evolution of the true orbital elements $e_0$
  as well as the corresponding $e_0^{\ast}$ and $e_0^{\ast\ast}$
  for a triple system with
  $m_1 = m_2 = 10 \solarmass$, $m_3 = 10^4 \solarmass$,
  with $a_3 \approx 43.298\,a_0$ and
  $\iota_{\rm init} =110^{\circ}$ under Newtonian gravity.
Further details of the orbital parameters are listed in the caption of the figure.
The oscillation of $e_0$ on the inner orbital time scale is smoothed out
  for $e_0^{\ast}$ as a consequence of the first canonical transformation,
  and the oscillation of $e_0^{\ast}$ on the outer orbital time scale
  is smoothed out due to the second canonical transformation.

In general, the oscillation of the orbital elements
  on the outer orbital time scale tend to be more
  pronounced compared to those on the inner orbital time scale.
This observation is affirmed by the fact that
  $S^{\ast}_{(i),n,m/2}$ is typically larger in magnitude than
  $S_{(i),n,m/2}$ for the same $(n,m/2)$ order,
  as indicated by their respective scaling with $a_0$ and $a_3$ (Eq.~\ref{eq:SDef} and Eq.~\ref{eq:SastDef}).
As depicted in Fig.~\ref{fig:OscillatingComponents},
  the discrepancy between $e_0^{\ast}$ and $e_0$ arising
  from the first canonical transformation
  varies on both the inner orbital time scale and outer orbital time scale.
During this transformation,
  the Hamiltonian is averaged over the inner orbital period
  while assuming that outer binary remains static.
This scenario is akin to a binary moving within
  a stationary external tidal field.
The distortion is dependent upon both the orbital phases
  of the inner and outer binaries.
On the inner orbital time scale,
  given that $e_0^{\ast}$ characterises the orbit's time-averaged properties,
  the maximal deviation between $e_0^{\ast}$ and $e_0$
  occurs near the periapsis and apoapsis of the inner binary.
This alignment is particularly evident in the early stages ($t \le 0.1\,{\rm yr}$).
On the outer orbital time scale, the
  deviation reaches the maximum at the periapsis of the outer binary,
  aligning with the strongest tidal force.
As shown in the second panel of Fig.~\ref{fig:OscillatingComponents},
  the instance of the maximum deviation ($e_0-e_0^{\ast}$) coincides
  with the projection of the tidal force onto the inner orbital plane
  (i.e. $\propto |(\hat{R_0} \times \hat{P_0}) \times \hat{R_3}|/R_3^3$).
The oscillation diminishes as eccentricity peaks
  around $t \sim 0.25\,{\rm yr}$,
  when the primary influence of the tidal force shifts to the precession
  of the inner orbit.

The deviation of $e_0^{\ast\ast}$ from $e_0^{\ast}$ arises
  from the second canonical transformation,
  signifying the deviation of the inner orbit
  from a Keplerian orbit due to the varying
  gravitational pull of the third body.
As expected, this deviation shows a rough
  periodicity aligned with the orbital time scale of the outer binary,
  peaking around the periapsis with a slight offset
  due to the relative inclination alignment of two binary orbits.
As illustrated in Fig.~\ref{fig:OscillatingComponents},
  a clear synchronization between the component of the tidal force perpendicular to the inner orbital plane
  (i.e., $\propto (\hat{R_0} \times \hat{P_0}) \cdot \hat{R_3}/R_3^3$)
  and eccentricity oscillation
  (i.e., $e^{\ast}_0 - e^{\ast\ast}_0$) can be observed.
When this component of the tidal force reaches its maximum,
  the perturber $m_3$ is nearly directly above or below the inner orbital plane.
The projection of the force exerted by the perturber onto
  each object in the inner binary
  directs inwards, forcing the binary
  to get closer and distorting it from a Keplerian orbit.
This configuration leads to the maximum deviation of $e_0^{\ast\ast}$ from $e_0^{\ast}$.

\begin{figure*}[htb]
\centering
\includegraphics[width=1.8\columnwidth]{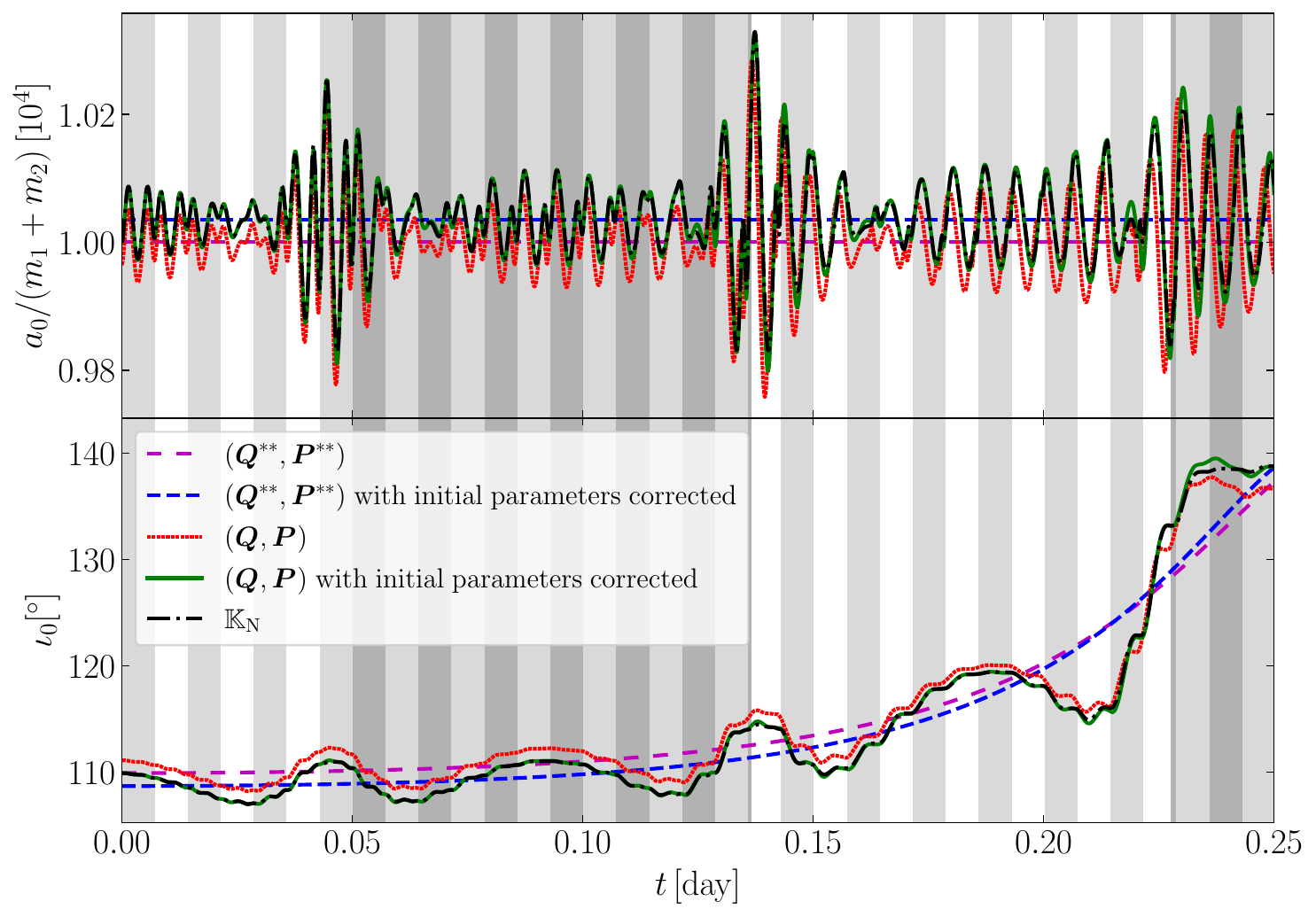}
\caption{
\label{fig:InitialCondition}
The triple system is composed of $m_1 = 10 \solarmass$,
   $m_2 = 10 \solarmass$
  with a $m_3 = 10^4 \solarmass$ perturber.
The orbital parameters for black-dashdotted,
  blue-dashed and green-solid lines are the same
  as the one in Fig.~\ref{fig:OscillatingComponents},
  whereas the orbital parameters for the rest are similar.
The loosely dashed magenta line and dotted red line represent the triple system
  with orbital parameters $a_0^{\ast\ast} = 10^4 (m_1 + m_2)$,
  $\iota^{\ast\ast} = 110^{\circ}$,
  $a_3^{\ast\ast} = 43.298 a_0^{\ast\ast}$,
  $\omega_0^{\ast\ast} = \pi/2$, $\omega_3^{\ast\ast} = 0.1$,
  $e_0^{\ast\ast} = 0.2$, $e_3^{\ast\ast} = 0.6$,
  $l_0^{\ast\ast} = 0$ and $l_3^{\ast\ast} = \pi$.
The densely dashed blue line and solid green line
  represent the triple system with initial condition corrected.
The DA orbital parameters are
  $a_0^{\ast\ast} \approx 1.004 \times 10^4 (m_1 + m_2)$,
  $a_3^{\ast\ast} \approx 43.14 a_0^{\ast\ast}$,
  $\iota^{\ast\ast} \approx 108.78^{\circ}$,
  $\omega_0^{\ast\ast} \approx 1.576$,
  $\omega_3^{\ast\ast} \approx 0.098$,
  $e_0^{\ast\ast} \approx 0.2023$,
  $e_3^{\ast\ast} \approx 0.2000$, $l_0^{\ast\ast} \approx 0.003375$
  and $l_3^{\ast\ast} \approx \pi$.
This set of DA orbital parameters corresponds to the same
  contact elements as in Fig.~\ref{fig:OscillatingComponents}.
The DA Hamiltonian used is
  $K^{\ast\ast} = K^{\ast\ast}_{(0),0,3} + K^{\ast\ast}_{(0),0,4}
  + K^{\ast\ast}_{(1),0,9/2}$.
The dashdotted black line represent the result from N-body simulation.
}
\end{figure*}

The conversion between orbital elements and DA elements
  is also crucial for recovering the correct orbital elements
  necessary for precise and reliable prediction of the system's evolution.
Moreover, it plays a key role in constructing
  the correct initial conditions \cite{Luo2016}.

In Fig.~\ref{fig:InitialCondition}, we
  demonstrate the impact of subtle variations in initial conditions on the system's evolution,
Figure~\ref{fig:InitialCondition} presents results
  from secular perturbation theory under different initial conditions,
  alongside those from N-body simulation for comparison.
Specifically, we compare results with initial conditions
  $a_0 = 10^4 (m_1 + m_2)$, $\iota = 110^{\circ}$
  and $e_0 = 0.2$, with those from a naive condition
  where $a_0^{\ast\ast}$, $\iota^{\ast\ast}$, and $e_0^{\ast\ast}$ are set to their corresponding values instead.
In the latter case, although the recovered orbital elements exhibit similar oscillatory behavior
  compared to the N-body simulation results,
  an apparent offset is observed due to this discrepancy in initial data.
Additionally, inclination oscillation is underestimated.
Conversely, when the initial conditions are corrected using
  the generating function,
  the DA orbital elements become
  $a_0^{\ast\ast} \approx 1.004 \times 10^3 (m_1 + m_2)$,
  $e_0^{\ast\ast} \approx 0.2023$ and $\iota^{\ast\ast} \approx 108.78^{\circ}$.
As shown in Fig.~\ref{fig:InitialCondition}, the evolution of the triple system predicted by the
  \ac{da} Hamiltonian aligns closely with predictions from N-body simulation.
However, a small offset exists between these two results,
  which becomes pronounced at $t \sim 0.25$ days.
This discrepancy may arise from higher-order interaction terms
  or could be related to the fundamental limitations of
  secular perturbation theory  (see Sec.~\ref{sec:Limitation} for detailed discussion).

As evidenced by the periodic oscillations of $a_0$ and $a_3$,
  periodic energy exchange occurs between the two binaries \citep{Luo2016,Kuntz2023}.
The absence of two mean anomalies $l_0^{\ast\ast}$ and $l_3^{\ast\ast}$
  from the final DA Hamiltonian
  implies that both semi-major axes $a_0^{\ast\ast}$ and $a_3^{\ast\ast}$
  remain constant, and hence the absence of secular energy exchange.
This observation aligns with the standard picture of \ac{vzlk} oscillations.
However, such adiabatic evolution does not hold near resonances of mean motions or those involving mean motion
  \citep[see, e.g.,][]{Kuntz2022}.
We address this topic in Sec.~\ref{sec:Limitation}.

\section{\label{sec:results}Results}

\begin{figure}[t]
\centering
\includegraphics[width=1\columnwidth]{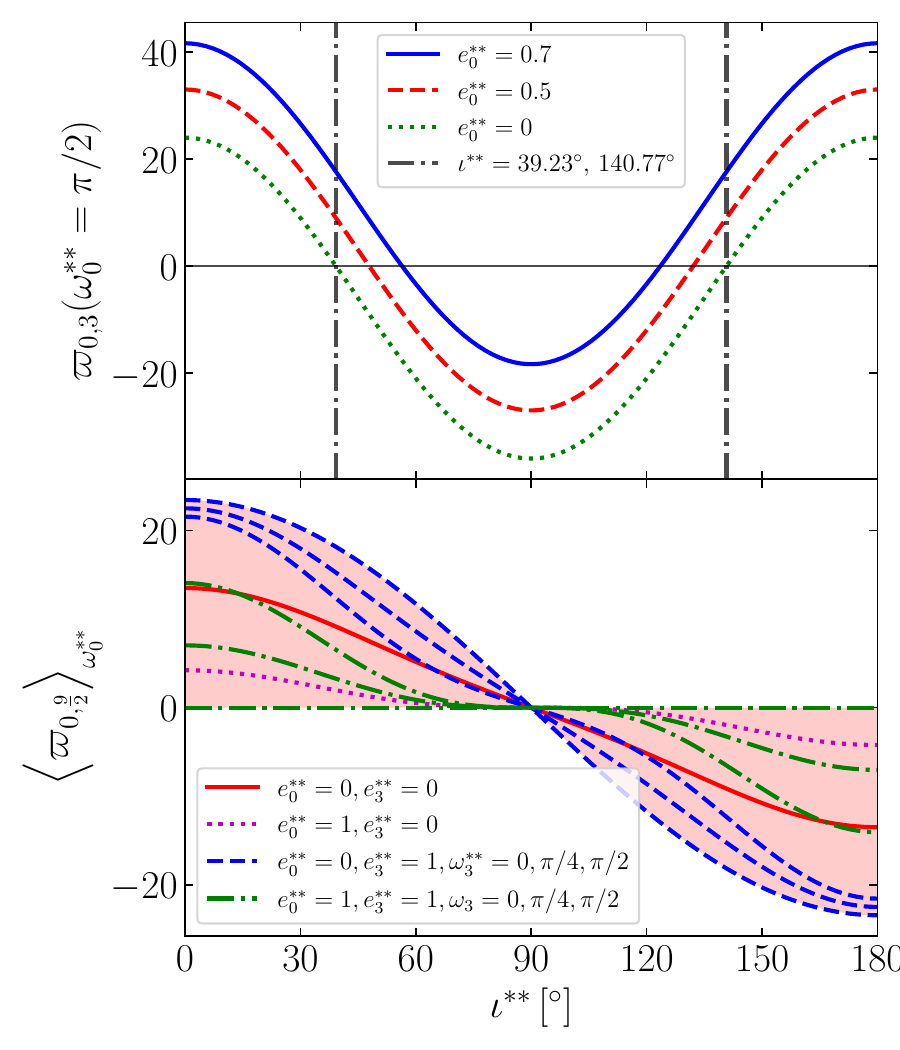}
\vspace*{-0.8cm}
\caption{
\label{fig:C09}
The figure shows the selected and averaged precession rates resulting from the quadrupole
  term (top panel) and $K^{\ast\ast}_{(1),0,9/2}$ term (bottom panel).
In the top panel, three curves correspond to initial eccentricities
  $e^{\ast\ast}_{0} = 0.7$ (blue-solid), $0.5$ (red-dashed) and $0$ (green-dotted),
  from top to bottom.
Vertical lines at $\iota^{\ast\ast} = 39.23^{\circ}$ and $140.77^{\circ}$
  indicate where the $e^{\ast\ast}_{0} = 0$ curve intersects zero,
  signifying the critical values between the libration island and circulation.
In the bottom panel, the precession rate is averaged over
  the argument of periapsis $\omega^{\ast\ast}_0$,
  resulting in consistently positive precession rates
  for prograde configurations and negative precession rates for retrograde configurations.}
\end{figure}

Quadrupole interaction plays a dominant role in the long-term
  evolution of the triple system, with higher-order terms serving as corrections
  to the quadrupole picture.
At the quadrupole level, the angular momentum $J_3^{\ast\ast} \gg J_0^{\ast\ast}$,
  indicating that $\iota^{\ast\ast}_3 \ll \iota^{\ast\ast}$ for b-EMRI systems.
If we adopt the assumption that $\iota^{\ast\ast}_3=0$ throughout the system's evolution
  \citep[i.e., the the test-particle-quadrupole approximation, see, e.g.,][]{Katz2011,Lithwick2011}
  as $\partial K^{\ast\ast}_{(0),0,3}/\partial \delta \Omega^{\ast\ast} \propto \sin \iota^{\ast\ast}_3$,
  $H^{\ast\ast}_0$ becomes a constant of motion. 
This conservation, combined with energy conservation,
  suggests that the system evolves along a closed loop on the
  $(\cos \omega^{\ast\ast}_0, e^{\ast\ast}_0)$ diagram.
This observation has led to the tradition of describing the system's evolution
  in this diagram, commonly referred to as a phase space diagram.
For an initial orbital configuration of $\omega^{\ast\ast}_0 = \pi/2$ and $e^{\ast\ast}_0=0$,
  the loop resembles a circulation curve at low inclination angles
  ($\iota^{\ast\ast}_{\rm init} \lessapprox 39.23^{\circ}$ or $\iota^{\ast\ast}_{\rm init} \gtrapprox 140.77^{\circ}$),
  with negligible variation in $e^{\ast\ast}_0$ (see, e.g., Fig.~\ref{fig:C09}).
When the inclination surpasses the critical value,
  libration islands form, allowing the eccentricity to grow,
  reaching a maximum value \citep{Holman1997}
\begin{align}
\label{eq:maxe0}
\max e_0^{\ast\ast}  \simeq \sqrt{1 - \frac{5}{3} \cos^2 \iota_{\rm init}} \ ,
\end{align}
for an initially circular orbit.
The oscillation time scale reads
\begin{align}
\label{eq:tQuad}
t_{\rm quad}  = 2 \pi \frac{\sqrt{M_0}}{m_3} \frac{{a_3}^3}{{a_0}^{3/2}} \left( 1 - {e_3}^2 \right)^{3/2} \ .
\end{align}
%
%
Interaction and relativistic terms beyond the quadrupole order
  affect the evolution of the system
  via perturbing the quadrupole oscillation mainly by accelerating or decelerating
  the precession rate.
We have
\begin{widetext}
\begin{align}
\label{eq:omegaDot}
\dot{\omega}_0^{\ast\ast} \big|_{1,0}
  & =
\sqrt{\frac{M_0}{a^{\ast\ast}_0{}^3}}
\frac{M_0}{a^{\ast\ast}_0}
\frac{3}{1-e^{\ast\ast}_0{}^2} \ ,
\nonumber \\
\dot{\omega}_0^{\ast\ast} \big|_{0,3}
  & =
    \varpi_{0,3} \sqrt{\frac{M_0}{a^{\ast\ast}_0{}^3}}
       \left(\frac{a^{\ast\ast}_0}{a^{\ast\ast}_3}
       \right)^3 \frac{m_3}{M_0}
    \frac{1}{(1-e^{\ast\ast}_0{}^2)^{1/2}(1-e^{\ast\ast}_3{}^2)^{3/2}}
    \ ,
    \nonumber \\
\dot{\omega}_0^{\ast\ast} \big|_{0,4}
  & =
  \varpi_{0,4} \sqrt{\frac{M_0}{a^{\ast\ast}_0{}^3}}
        \left(\frac{a^{\ast\ast}_0}{a^{\ast\ast}_3}  \right)^4
    \frac{\Delta m}{M_0} \frac{m_3}{M_0} \frac{e^{\ast\ast}_3}{e^{\ast\ast}_0 (1-e^{\ast\ast}_0{}^2)^{1/2}
    (1-e^{\ast\ast}_3{}^2)^{5/2}}  \ ,
    \nonumber \\
\dot{\omega}_0^{\ast\ast} \big|_{0,\frac{9}{2}}
  & =
   \varpi_{0,\frac{9}{2}} \sqrt{\frac{M_0}{a^{\ast\ast}_0{}^3}}
     \left(\frac{a^{\ast\ast}_0}{a^{\ast\ast}_3}  \right)^{9/2}
       \left( \frac{m_3}{M_0} \right)^{3/2}
       \frac{\sqrt{m_3}}{\sqrt{M_3}}
       \frac{1}{\left(1-{e^{\ast\ast}_3}^2\right)^3}  \ .
\end{align}
\end{widetext}
where $\dot{\omega}_0^{\ast\ast} \big|_{1,0}$
  is the  1PN precession term.
The symbol $\varpi_{n,m/2}$ above denotes the component of precession rate
  that is a function of trigonometric functions
  of angles, which exhibits weak dependence on both eccentricities.
The precession rate due to quadrupole and cross-term contributions
  is more significant for a massive third body on a highly eccentric orbit.
This suggests that eccentricity excitation is particularly relevant
  for b-EMRI systems, where a binary is gravitationally captured by a supermassive black hole.

The relativistic correction $K^{\ast\ast}_{(0),1,0}$ consistently results in a positive precession rate,
  irrespective of the relative inclination angle.
This correction disrupts the quadrupole libration islands, forcing them to circulate instead,
  thereby suppressing the eccentricity excitation \citep{Ford2000,Naoz2013b}.
Motivated by these observations, \citep{Naoz2013b} introduced
\begin{align}
\Lambda = \frac{1}{3 \delta}
   \frac{\alpha^3 q_m }{(1-{e_3}^2)^{3/2}}
   \sim \frac{\dot{\omega}^{\ast\ast}_{0}|_{0,3}}{\dot{\omega}^{\ast\ast}_{0}|_{1,0}} \
\end{align}
to gauge the significance of the relativistic interaction.
When $\Lambda \gg 1$, the system behaves similarly to classical \ac{vzlk} oscillations,
  where Newtonian gravity dominates.
Conversely, when $\Lambda \ll 1$, the relativistic interaction
  dampens eccentricity oscillation, effectively decoupling the evolution of the inner binary from the perturber.

\begin{figure*}[htb]
\centering
\includegraphics[width=2.0\columnwidth]{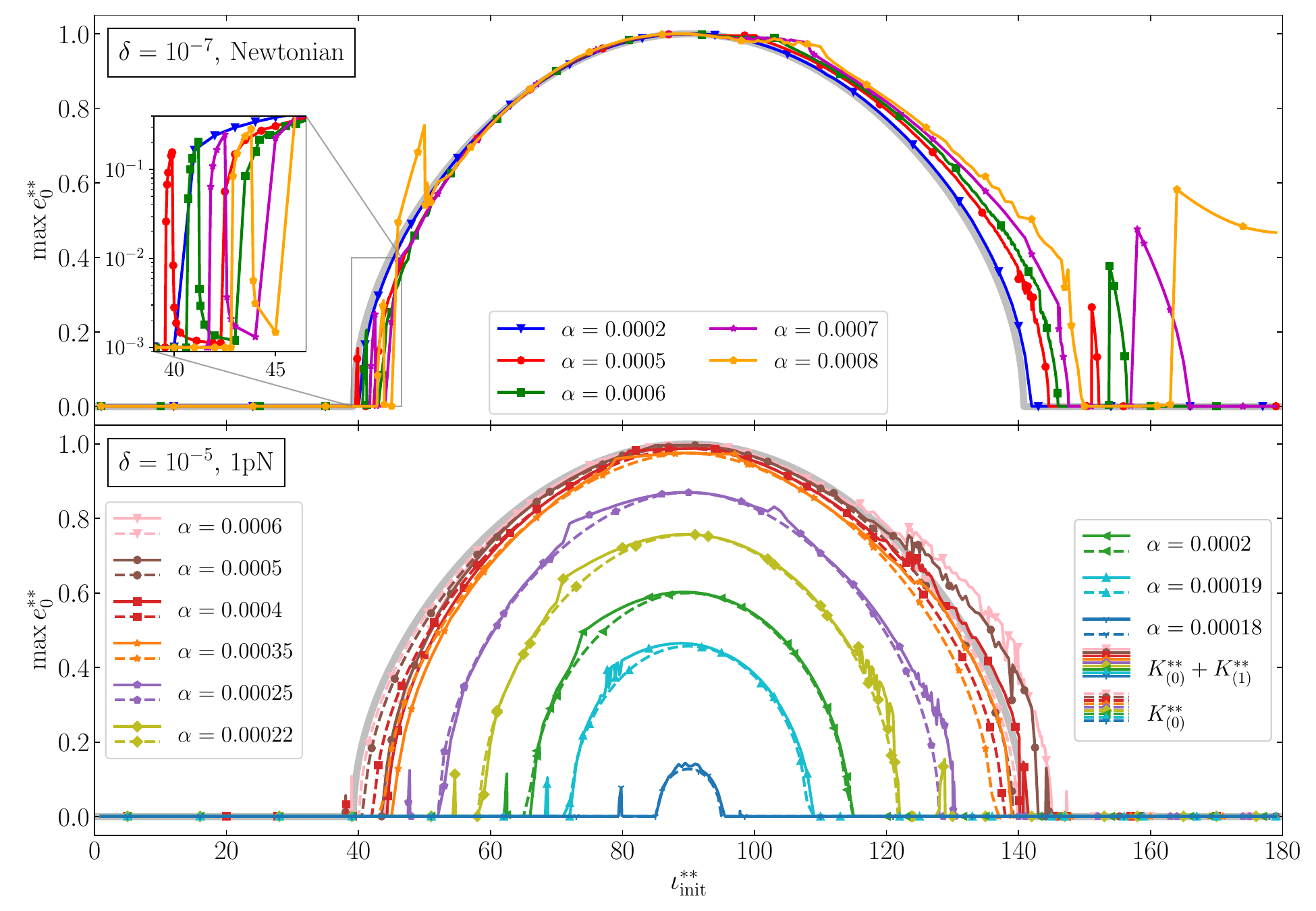}
\caption{
\label{fig:Newtonian_pN}
The figure displays the maximum eccentricity
  for several systems evolved using Newtonian DA Hamiltonian (top panel)
  and relativistic DA Hamiltonian (bottom panel).
All systems have $m_1=m_2 = 20\solarmass$, $m_3 = 2\times 10^7 \solarmass$,
  $a_0 = M_0/\delta$, $a_3 = a_0/\alpha$ and
  $\omega_0^{\ast\ast} = 90^{\circ}$, $\omega_3^{\ast\ast} = 270^{\circ}$,
  $e_0^{\ast\ast} = 10^{-3}$, $e_3^{\ast\ast}=0.8$.
Each system is evolved for  for $500 \,t_{\rm quad}$.
In the top panel, the colored lines with markers
  represent predictions from $K^{\ast\ast}_{(0),0,3} + K^{\ast\ast}_{(1),0,9/2}$.
The thick gray line shows the prediction from
  $K^{\ast\ast}_{(0),0,3}$ (i.e., the classical \ac{vzlk} resonance),
  which is independent of the value of $\alpha$.
The narrow resonance peaks arise from the interaction between
  the quadrupole-squared term and the quadrupole term
  (top panel) or their complex interplay with the relativistic terms.
}
\end{figure*}

The contributions of other interaction terms are in general inclination dependent.
Following a similar philosophy, we define a new parameter
\begin{align}
\Upsilon = \frac{\alpha^{3/2} {q_m}^{1/2}}{(1-{e_3}^2)^{3/2}}
   \sim \frac{\dot{\omega}^{\ast\ast}_{0}|_{0,9/2}}{\dot{\omega}^{\ast\ast}_{0}|_{0,3}} \
\label{eq:Ysilon}
\end{align}
to describe the contribution of the $K^{\ast\ast}_{(1),0,9/2}$ cross term.
As depicted in Fig.~\ref{fig:C09}, the averaged $\varpi_{0,9/2}$
  remains positive for $\iota < 90^{\circ}$ and negative
  for $\iota > 90^{\circ}$, regardless of the other orbital parameters.
Consequently, this cross term shifts the critical inclination angles
  between circulation and libration islands to larger values, resulting in a shift
  of the range of orbital parameters exhibiting eccentricity excitation,
  as shown in the upper panel of Fig.~\ref{fig:Newtonian_pN}.
The systems with $\iota_{\rm init} \sim 40^{\circ}$ are now situated within the circulation domain,
  while systems with $\iota_{\rm init} \sim 140^{\circ}$ exhibit
  eccentricity excitation that are otherwise absent in quadrupole-level approximation.
This phenomenon is reminiscent of the non-test-particle effect studied by \citep{Hamers2021}.

In Fig.~\ref{fig:Newtonian_pN}, we demonstrate the
  eccentricity oscillations for b-EMRI systems with different $\alpha$ values and inclination angles.
We compare the results obtained from Newtonian and relativistic triple systems
  using secular perturbation theory, both with and without cross-term effects.
In the top panel, we demonstrate eccentricity excitation under Newtonian gravity.
For systems with $\Delta m = 0$, the only secular terms are the quadrupole term and the
  quadrupole-squared term $K^{\ast\ast}_{(1),0,9/2}$.
The system governed solely by quadrupole perturbation follows the simple relation in
  Eq.~\ref{eq:maxe0},
  as shown by the thick grey curve in the top panel.
It is evident that $K^{\ast\ast}_{(1),0,9/2}$ is solely responsible for many of
  the resonance peaks across the extensive parameter space.
We identify a resonance peak near $\iota_{\rm init}=40^{\circ}$ and another near $150^{\circ}$,
  both of similar origin.
Additional resonance peaks near $50^{\circ}$ and $110^{\circ}$,
  along with several unresolved ones between $120^{\circ}$ and $145^{\circ}$,
  are observed when the third body is sufficiently close
  (i.e., $\alpha \ge 0.0008$ for this set of parameters).
These resonance peaks persist in the relativistic cases,
  indicating that the quadrupole-squared cross term continues to play a significant role in the dynamics of relativistic triple systems.
The analysis in relativistic cases is more complex
  due to the presence of multiple time scales.
If the third body's tidal force cannot compete with
  the short-range relativistic force,
  the resonance region shrinks,
  and the maximum eccentricity is reduced.
This has been observed in studies of relativistic effects,
  tidal interactions, and rotation \citep[e.g.,][]{Wu2003,Fabrycky2007,Naoz2013b,Liu2015}.
In the weakly relativistic regime, the interplay of different interaction terms
  adds complexity to the system's evolution,
  shifting the location of Newtonian resonance peaks and
  allows for the possibility of resonances involving three or more interaction terms.
When the system is dominated by quadrupole contribution (e.g., the $\delta = 10^{-5}$, $\alpha=0.0006$ case),
  it behaves similarly to the Newtonian case, with some Newtonian resonance peaks
  suppressed by relativistic precession (e.g., those near $40^{\circ}$ and $150^{\circ}$)
  and others further excited (e.g., those between $120^{\circ}$ and $145^{\circ}$).

While these resonance peaks are intriguing, not all are necessarily physical,
  as some occur when secular perturbation theory breaks down.
Due to computational limitations, a comprehensive investigation of
  these resonance peaks is deferred to future studies.
In the following subsections we identify several resonance peaks
  as case studies to examine the interplay of the quadrupole-squared cross term
  with other direct terms.
The quadrupole-squared term has been shown to improve the agreement between
  secular perturbation theory and N-body simulations in various studies
  \citep[e.g.,][despite differences due to different choices of fictitious time]{Luo2016,Will2021,Tremaine2023}.
Having confirmed the exact agreement in evolution equation with \citep{Will2021}
  and hence several others \citep{Luo2016,Kuntz2023,Tremaine2023},
  we find it unnecessary to repeat their case studies to showcase
  the power of this term to improve the agreement between secular perturbation
  and N-body simulation results.
We instead shift focus to marginal cases where cross-term effects do
  reproduce N-body simulation results (albeit with subtle differences) and
  aim to fully explore the capabilities and limitations of secular perturbation theory.
In Sec.~\ref{sec:CaseNewtonian}, we discuss Newtonian resonance,
  focusing on the resonance peaks
  near $40^{\circ}$ and $150^{\circ}$.
In Sec.~\ref{sec:CaseRelativistic}, we discuss relativistic resonance,
  focusing on an interesting case exhibiting chaotic behavior.

\subsection{Case study: Newtonian triple system} \label{sec:CaseNewtonian}

\begin{figure*}[htb]
\centering
\includegraphics[width=2\columnwidth]{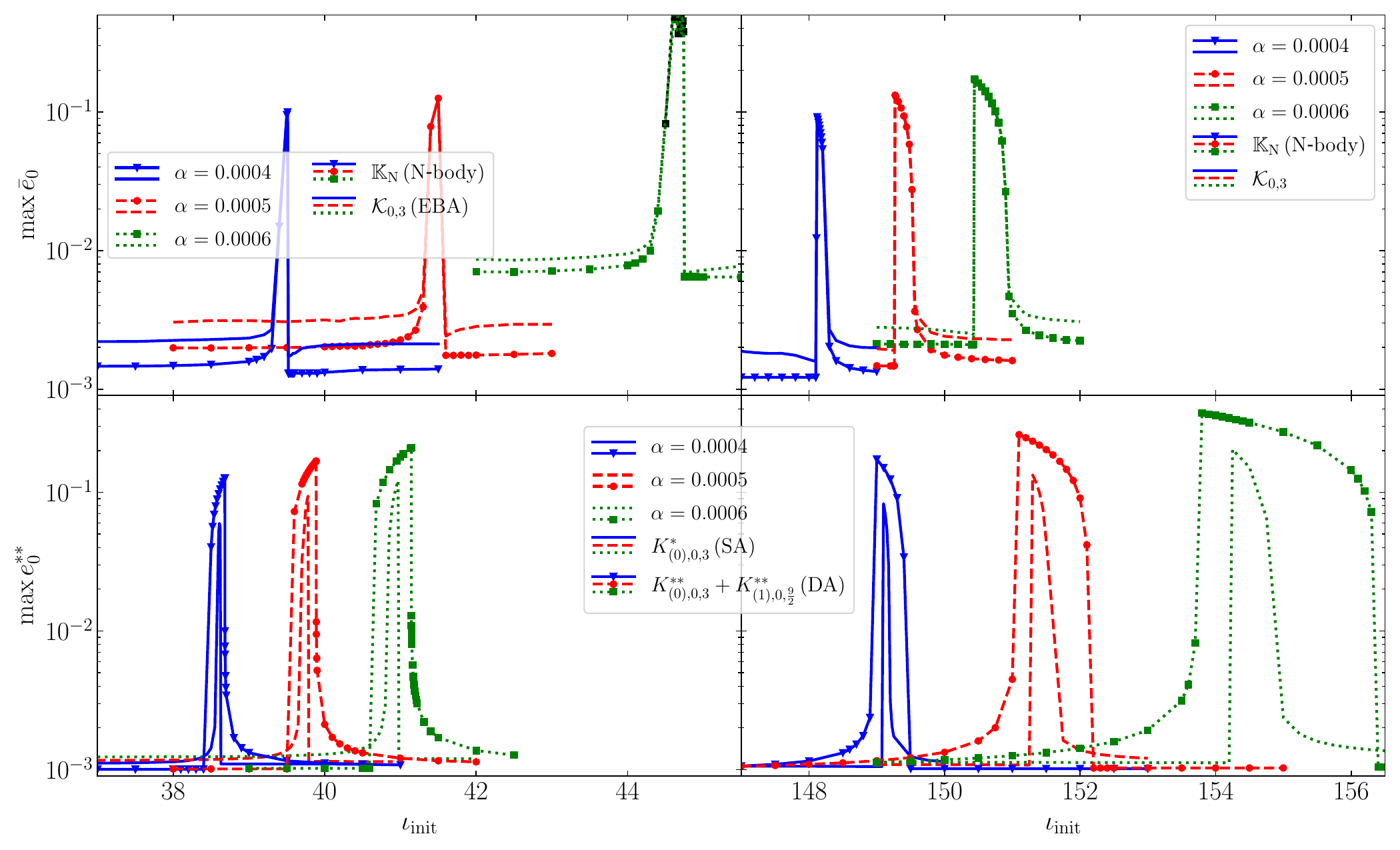}
\vspace*{-0.5cm}
\caption{
\label{fig:KastastK3Newtonian}
This figure depicts a zoomed-in view of the resonance peaks near
  $\iota_{\rm init} = 40^{\circ}$ and $150^{\circ}$
  in both N-body simulations (top two panels)
  and secular perturbation theory (bottom two panels).
In the top two panels, the mean eccentricities $\bar{e}_0$
  are calculated by a low-pass filter with a cutoff frequency
  of $1/(2P_3)$ to remove the oscillations at orbital periods.
Systems with variations in $\bar{a}_0$
  exceeding $10^{-3}a_0$ are marked with black squares.
The systems have $m_1 = m_2 = 20 \, M_{\odot}$, $m_3 = 2 \times 10^7 \, M_{\odot}$,
  $a_0 = 10^7 M_0$,
  $e_0 = 10^{-3}$, $e_3 = 0.8$, $\omega_0 = 90^{\circ}$, and $\omega_3 = 270^{\circ}$ initially.
All systems here are evolved for $500 \, t_{\rm quad}$.
Secular perturbation theory predicts a resonance peak similar to
  that observed in N-body simulations,
  albeit with discrepancies in its location and width.}
\end{figure*}

\begin{figure*}[htb]
\centering
\includegraphics[width=2\columnwidth]{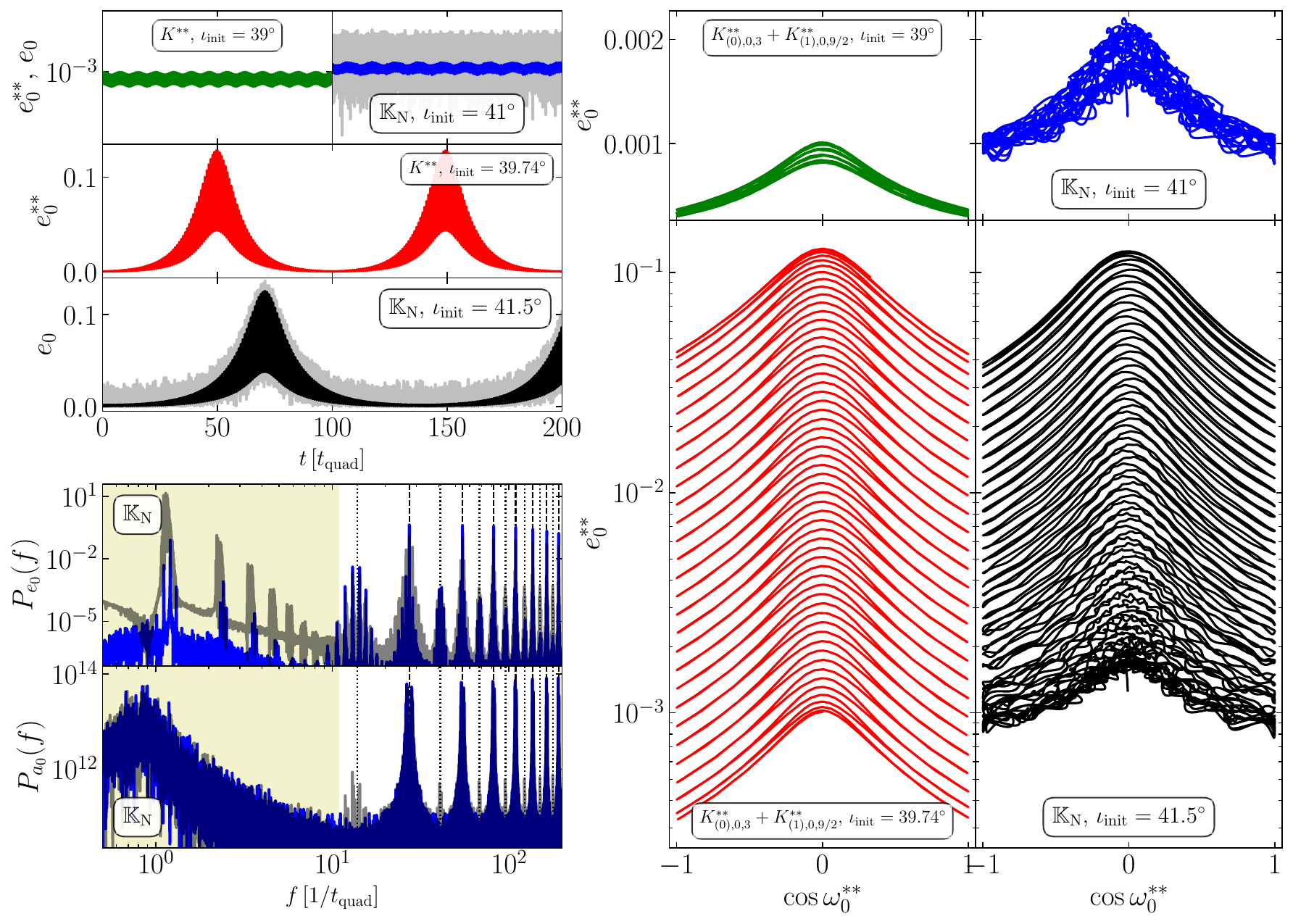}
\vspace*{-0.4cm}
\caption{
\label{fig:resonance}
This figure presents the evolution of several selected systems
  from Fig.~\ref{fig:KastastK3Newtonian}.
The initial inclination angles chosen for the N-body simulations
  are $41^{\circ}$ (non-resonant, blue) and
  $41.5^{\circ}$ (at resonance, black),
  while for the DA Hamiltonian,
  the angles are $39^{\circ}$ (non-resonant, green) and
  $39.74^{\circ}$ (at resonance, red).
In the top-left and top-right sub-panels of the
  N-body simulation results,
  the colored lines represent the mean eccentricities $\bar{e}_0$,
  calculated using a low-pass filter with
  a cutoff frequency of $1/(2P_3)$, while the grey lines depict $e_0$
The two resonance cases exhibit similar evolutionary patterns
  in phase space, particularly for $e^{\ast\ast}_0 \gtrsim 10^{-2}$.
In contrast, for $e^{\ast\ast}_0 \lesssim 10^{-2}$,
  the eccentricity evolution is dominated by rapid oscillations
  occurring at the frequency of $(1:m)$ orbital resonance
  (indicated by the vertical dashed lines in the bottom-left panels; see text for details).
This behavior is especially pronounced in the non-resonance cases.
}
\end{figure*}

Near $\iota^{\ast\ast}_{\rm init} \sim 40^{\circ}$ and $\sim 150^{\circ}$, the
  transition regime between circulation and libration regime of classical quadrupole perturbation,
  there are narrow resonance peaks that appear in both Newtonian triple systems
  and persist in the weakly-relativistic systems.
In Fig.~\ref{fig:KastastK3Newtonian} and Fig.~\ref{fig:resonance},
  we show the behavior of Newtonian triple systems predicted by the secular perturbation
  theory, compared against the N-body simulation.

Figure~\ref{fig:KastastK3Newtonian} present the evolution of several systems
  with parameters $m_1 = m_2 = 20 \solarmass$, $m_3 = 2\times 10^7 \solarmass$,
  $e_3 = 0.8$, $e_0 = 10^{-3}$, $\omega_0 = 90^{\circ}$ and $\omega_3 = 270^{\circ}$
  with various inclination angles and $\alpha = 0.0004,0.0005,0.0006$
  until $500 \,t_{\rm quad}$.
The figure shows the maximum eccentricity of each system during this period.
We also include the results from SA Hamiltonian and \ac{eba} for comparison.
Secular perturbation theory predicts resonance peaks
  akin to those in the N-body simulation,
  despite a small mismatch in their locations and width at the $\sim 2^{\circ}$ level.
These peaks originate from the same physical mechanism,
  as evidenced by the consistency in their location with respect to $\alpha$ values,
  as well as their similar phase space evolution shown in Fig.~\ref{fig:resonance}.
Such narrow resonance peaks indicate a precise matching between
  different components of interaction terms,
  and are therefore very sensitive to the system parameters and input Hamiltonian.
The subtle difference in the ingredients used in secular perturbation theory
  and N-body simulation suggest that such resonance peaks might differ
  in shape, width and location.
Comparing the results from \ac{sa} and \ac{da}
  reveals that discrepancies in peak shape and width are likely
  due to the exclusion of higher-order quadrupole-squared term
  and quadrupole-cubic term, etc, in the second canonical transformation,
  whereas peak location is mainly influenced by those arising
  from the first canonical transformation.
Additionally,
 the $(1:m)$ resonances
  between two orbital frequencies (with $m$ being the nearest integer
  to $P_3/P_0$, see Sec.~\ref{sec:Limitation} for details) that are missing in the
  current secular perturbation framework also provide non-trivial contributions.

Figure~\ref{fig:resonance} shows the impact of these overlooked resonances.
This figure selects several cases simulated using
  \ac{da} and compares them against the N-body simulations
  that exhibit similar evolution patterns, examining the underlying dynamics
  of these resonance peaks.
For the \ac{da} system with $\iota_{\rm init} = 39^{\circ},39.74^{\circ}$,
  the precession rate $\dot{\omega}^{\ast\ast}_{0}|_{0,9/2}$ ($< 0$)
  is comparable to the precession rate
  $\dot{\omega}^{\ast\ast}_{0}|_{0,3}$ ($>0$).
Although the contribution from $(0,9/2)$ term is not strong enough to counterbalance
  the precession from the quadrupole term to drag the system out of
  circulation, the resonance between these two
  precession rate for system with $\iota_{\rm init} = 39.74^{\circ}$
  breaks the energy and angular momentum
  conservation at the quadrupole level.
Such accumulative drift in energy and angular momentum drives
  the system towards eccentricity excitation,
  allowing the system to move between different trajectories
  in phase space.
This behavior is particularly evident when comparing systems that
  exhibit resonance to those that do not (e.g., the $\iota_{\rm init} = 39^{\circ}$ case).
Similar patterns are observed in N-body simulations.
For instance, the $\iota_{\rm init} = 41.5^{\circ}$ case,
  simulated using N-body methods, exhibits a similar evolution pattern,
  particularly near eccentricity maxima,
  indicating that the resonances likely have the same physical origin.

In addition to the eccentricity excitation,
  small yet rapid oscillations of $e_0$ and $\omega_0$
  are observed in N-body simulations.
These oscillations have frequencies
\begin{align}
        f = \bigg| \frac{m \pm k}{P_3} - \frac{1}{P_0}
          \bigg|  \pm n \, ,
\end{align}
for $k,n=0,1,2,\dots$
 and $m = \nint[\big]{{P_3}/{P_0}}$,
 as shown by the power spectrum in the bottom-left panels.
%
The frequency suggests that these oscillations
  arise from orbital resonance between the two orbits,
  which is not captured by the current secular perturbation framework.
In light of such limitations, a meaningful comparison of
  the system evolution predicted by secular perturbation theory
  and that from N-body simulation near resonance
  should be interpreted with caution.
Any discrepancies observed could arise from higher-order direct and cross terms,
  which can be resolved by extending \ac{da} calculations to higher orders.
However, they could also stem from $(1:m)$ orbital resonances
  not accounted for in the current secular perturbation framework.
Although these resonances are typically associated with energy exchange,
  their effects become significant even before the adiabatic approximation breaks down.
As discussed in Sec.~\ref{sec:Limitation},
  the strength of this resonance scales with the orbital frequency
  at the periapsis of the outer binary,
  which is proportional to $\Upsilon$: the ratio of
  the quadrupole-squared precession rate to the quadrupole precession rate.
This alignment is not coincidental and originates
  from the tidal nature of quadrupole interactions
  and extends to the quadrupole-cubic term (see, e.g., \citep{Conway2024}).
Thus, identifying the parameter space where higher-order (quadrupole)
  cross terms
  are significant is equivalent to finding
  where $(1:m)$ resonances are important,
  and developing secular perturbation theory must account for both.
Neglecting these $(1:m)$ resonances could lead to
  overlooking essential elements necessary for
  accurately predicting the long-term evolution of triple systems.

\subsection{Case study: Relativistic triple system} \label{sec:CaseRelativistic}

\begin{figure*}[htb]
\centering
\includegraphics[width=2\columnwidth]{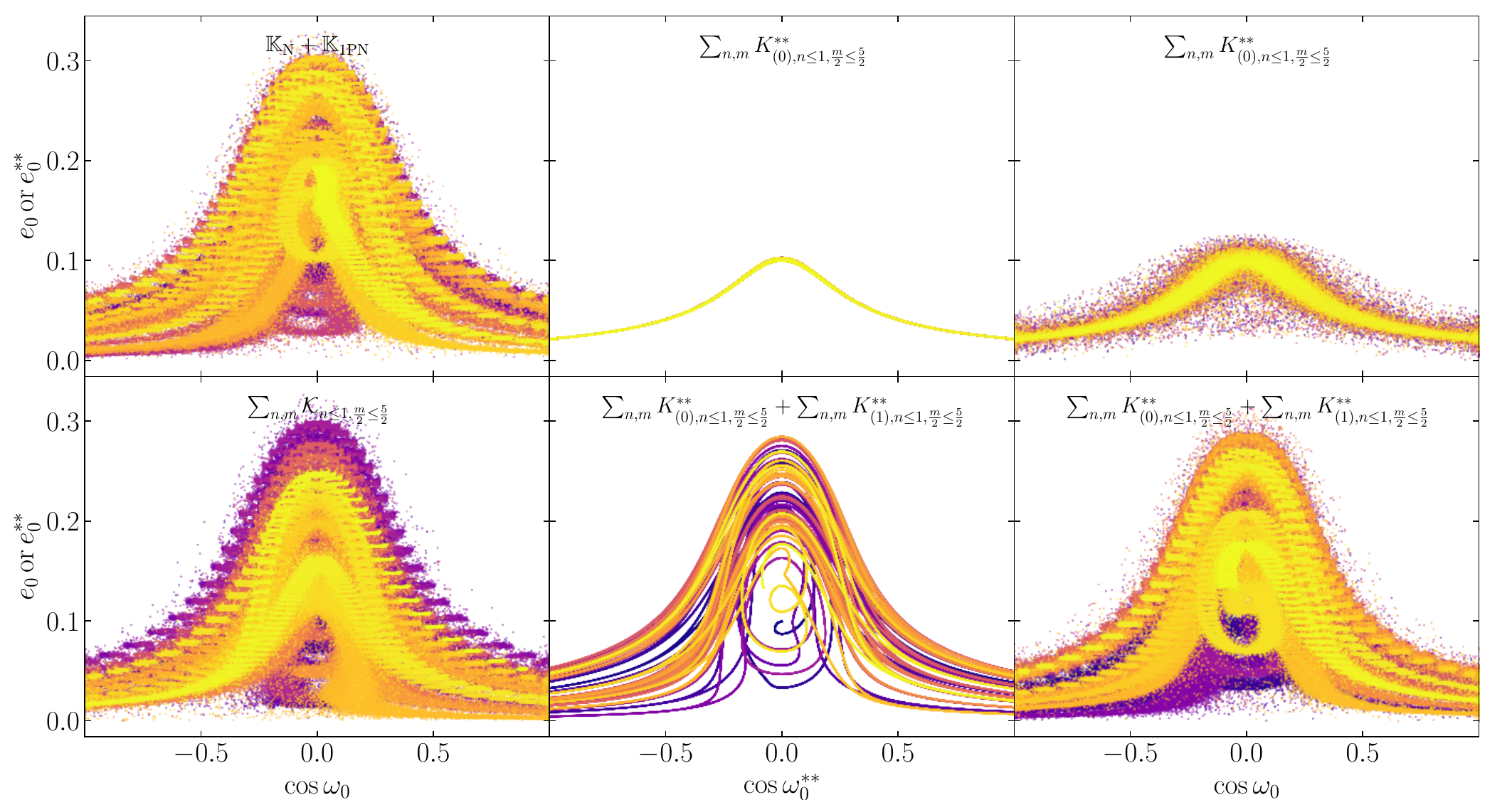}
\centering
\includegraphics[width=2\columnwidth]{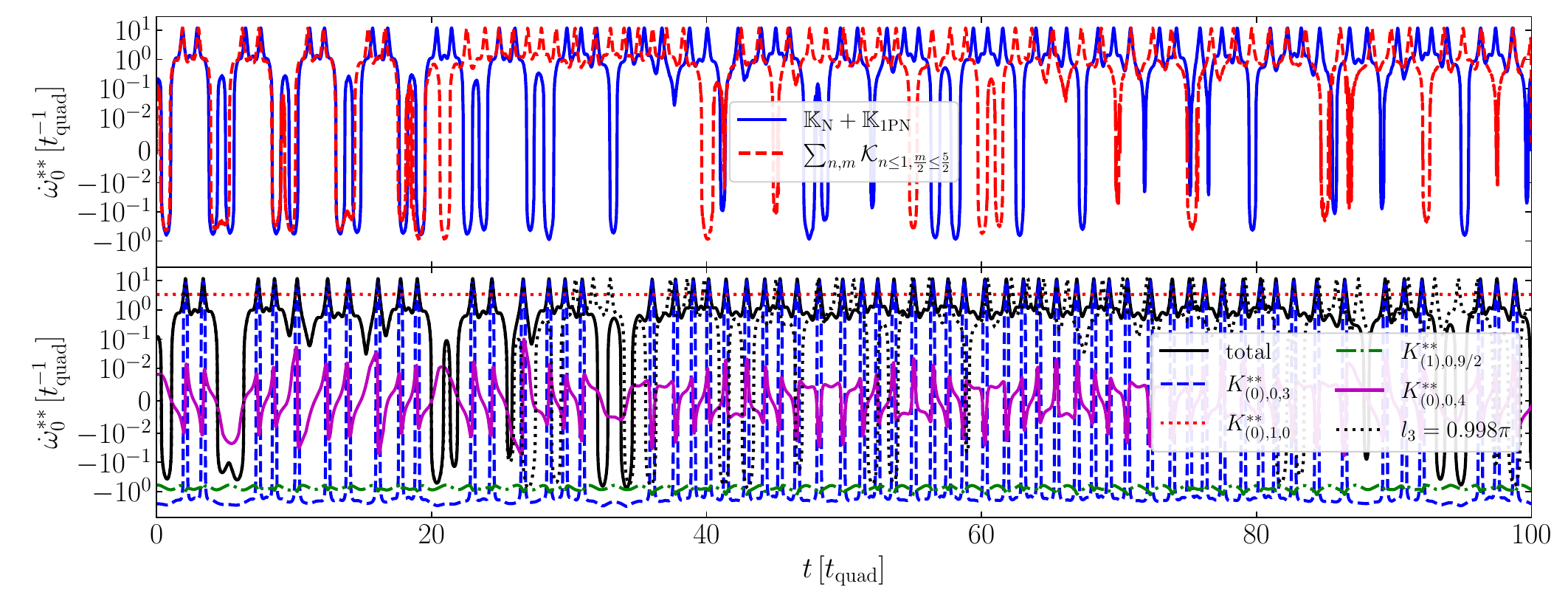}
\vspace*{-0.5cm}
\caption{
\label{fig:CrossDirectCompare}
The top six panels show the phase space evolution of a triple system with
  $m_1 = 20\solarmass$, $m_2 = 30\solarmass$, $m_3 = 2\times 10^7 \solarmass$,
  with initial $e_0 = 0.1$, $e_3 = 0.89$,
  $a_0 = 0.05 \,{\rm au}$, $a_3 = 210 \,{\rm au}$,
  $\omega_0 = \pi/2$, $\omega_3 = 280^{\circ}$,
  $\iota = 134^{\circ}$, $l_0 = 0$ and $l_3 = \pi$ over
  $100 \;\! T_{\rm quad}$.
This system has $\delta \approx 9.85 \times 10^{-6}$,
  $\alpha \approx 2.38 \times 10^{-4}$
  and $\Lambda = 1.93$ and $\Upsilon = 0.0245$.
The top-left (bottom-left) diagram shows the results from N-body simulations
  using the Hamiltonian $\mathbb{K}_{\rm N} + \mathbb{K}_{\rm 1PN}$
  ($\sum \mathcal{K}_{n\le1,\frac{m}{2}\le \frac{5}{2}}$).
The top-middle and top-right diagrams display the results using
  the DA Hamiltonian without cross terms, with top-middle showing the DA orbital elements
  and top-right showing the true orbital elements.
Similarly, the bottom-middle and bottom-right diagrams show the results from
  DA Hamiltonian, including the cross terms.
The color of the points represents time,
  changing from dark purple to yellow as time increases.
As shown in the figure, the suppression of eccentricity oscillations
  due to relativistic terms is revived by the cross term,
  leading to chaotic evolution of the system in phase space.
In the bottom two panels,
  the precession rates from the first few leading-order terms
  are shown to further illustrate the chaotic evolution.
Instances where the (total) precession rate crosses zero clearly
  mark the transitions between the libration and circulation phases.}
\end{figure*}

While the relativistic term generally suppresses the eccentricity oscillation,
  the presence of the cross term significantly alters the dynamics.
The term $K^{\ast\ast}_{(1),0,9/2}$ enhances the suppression effect
  in prograde configurations, while in retrograde configurations it partially counteracts the relativistic precession,
  releasing the system from suppression.
This results in the appearance of a resonant peak
  near the transition between the circulation and libration regimes,
  akin to what is observed in Newtonian cases.
However, in contrast to the Newtonian scenarios,
  these resonances exhibit much richer and more intriguing dynamics.

Figure~\ref{fig:CrossDirectCompare} presents a selected system (see caption for system parameters)
  that exhibit such dynamics
  and compares the predictions from N-body simulation (i.e., $\mathbb{K}_{\rm 1PN}$),
  and those from secular perturbation theory with and without cross terms.
Additionally, we include one simulation using $\sum \mathcal{K}_{n,m/2}$ for comparison,
  referred to as the \ac{eba}.
Without relativistic corrections, this system undergoes libration,
  with eccentricity excited to $\max e^{\ast\ast}_0 \approx 0.50$.
With $\Lambda = 1.93$, the eccentricity oscillation is
  expected to be suppressed when relativistic corrections are included.
Indeed, as shown in the  top-middle and top-right panels of Fig.~\ref{fig:CrossDirectCompare},
  the maximum eccentricity decreases to approximately
  $\max e^{\ast\ast}_0 \approx 0.10$, 
  and the system undergo circulation, when only direct terms are included
  in the DA Hamiltonian.
However, this suppression is compensated by the cross term $K^{\ast\ast}_{(1),0,9/2}$.
The maximum eccentricity of $\max e^{\ast\ast}_0 \approx 0.28$ 
  is predicted when the cross term
  effect is included, consistent with the N-body result.

For this system,
  the contribution to precession from the 1PN term and the $K^{\ast\ast}_{(1),0,9/2}$ term
  are comparable, differing by a factor of approximately $4$.
The cross term, along with the quadrupole term, can counteract
  the relativistic precession, enabling the system to transition into a libration state around
  $\cos \omega_0^{\ast\ast} = 0$,
  and then return to the circulation state after a brief entrapment period.
As shown by Fig.~\ref{fig:CrossDirectCompare}, these transitions
  between circulation and libration appear chaotic,
  with N-body simulations predicting different transition instances
  compared to \ac{eba} or secular perturbation theory.
Even within secular perturbation theory, the transition instances
  appear to depend sensitively on the initial conditions,
  as demonstrated by the divergence between systems
  with different initial conditions (i.e. $l_3 = \pi$ and $l_3 = 0.998\pi$) around $t \sim 26 \, t_{\rm quad}$,
  as shown in the bottom panel of Fig.~\ref{fig:CrossDirectCompare}.
The discrepancy between secular perturbation theory and N-body simulation
  appears on a much shorter time scale, indicating that the conversion
  between orbital elements and DA orbital elements likely plays a minor role
  in the discrepancies between secular perturbation theory and N-body simultion.

Conversely, differences in predictions between N-body simulations
  and \ac{eba} suggest that higher order multipole expansions play a more significant role.
While these higher order terms do not directly affect the appearance and
  size of eccentricity resonance,
  their otherwise negligible impacts manifest in chaotic behaviors.
The inherent chaotic nature of these systems highlights the difficulty
  in accurately predicting their long-term evolution.
Additionally, when evaluating the reliability of secular perturbation theory
  through comparisons with N-body simulations,
  caution must be exercised.
Recognising the chaotic dynamics involved, carefully selecting meaningful observables for reliable comparisons is essential
  to ensure a more thorough assessment of the predictive capabilities of secular perturbation theory.

\section{Discussion}
\subsection{Comments on osculating elements and contact elements}
\label{sec:CommentsLagrangian}

It is well known that there exists a gauge freedom in the choice of
  orbital elements to describe a perturbed celestial orbit.
This freedom concerns how we divide the binary's physical velocity
  between the orbital velocity within a fictitious Keplerian orbit
  and the changes in the Keplerian orbit itself.
A comprehensive discussion on this subject can be found in \cite{Efroimsky2005}.

Instantaneous orbital elements chosen such that the tangent of
  the corresponding conic equals the velocity
  are referred to as osculating elements.
The osculating elements represent
  the Keplerian orbit the binary would follow
  if the perturbing force were instantaneously removed.
They are typically derived through
  the Lagrangian perturbation method,
  where the Lagrangian constraints necessary for osculation are naturally imposed.

Conversely, the orbital elements obtained through the Hamiltonian method,
  known as contact elements, are non-osculating \citep{Brumberg1991}.
For contact elements, the tangent of the
  instantaneous conic aligns with momentum instead of velocity.
The fundamental difference between osculating and contact elements
  lies in the difference between velocity and momentum,
  which becomes non-trivial when the perturbing force is velocity-dependent.
In a hierarchical triple system,
  where the Newtonian coupling force between binaries
  depends solely on their relative positions,
  the osculating elements coincide exactly with the contact elements at this order.

For this reason specifically,
  the evolution equation of the orbital elements
  resulting from the cross term $K^{\ast\ast}_{(1),0,9/2}$
  in our research perfectly matches that of the osculating elements
  derived by \cite{Will2021}.
However, at relativistic orders, where the coupling between binaries
  becomes velocity-dependent,
  discrepancies arise in the evolution equations
  between osculating and contact elements.
Comparing the corresponding evolution equations
  poses a non-trivial task beyond the current scope of this work.

Although osculating elements are often favoured in the literature for their natural description of orbital perturbation,
  they might not always be the optimal choice,
  especially in the context of
  relativistic secular perturbation theory \citep{Kuntz2023}.
On the one hand, the Hamiltonian approach naturally
  offers control over energy conservation,
  which is crucial for tracking numerical errors and
  verifying the calculation of cross terms.
On the other hand, issues with angular momentum conservation
  and the invariant reference plane (see Sec~\ref{sec:Elimination})
  highlight subtleties in the DA procedure,
  underscored by the counter-intuitive offset
  between the orbital and DA orbital elements
  presented by $S^{\ast}_{(1),n,m/2}$.
It is unclear how these issues manifest in osculating elements.

Furthermore, when comparing theoretical predictions
  to observations,  the natural choice of orbital elements
  hinges on the particulars of the observation.
This was discussed, for instance, in the context of the gravito-magnetic clock effect \cite{Li2024}.
Nonetheless, deriving orbital evolution using both
  contact and osculating elements remains crucial,
  providing cross-validation for complex calculations,
  especially for higher-order cross terms beyond the capability of manual computation.

\subsection{Comments on $\delta \Omega = \pi$ and angular momentum conservation}
\label{sec:Elimination}

The assumption $\delta \Omega = \pi$ is applied either explicitly
  or implicitly throughout the calculation of DA method
  in both Hamiltonian and Lagrangian approaches,
  to simplify the form of the evolution equation.
The mishandling of this condition has lead to
  incorrect results in the literature.
As highlighted by \citep{Naoz2013a,Naoz2016},
  substituting $\delta \Omega = \pi$
  at the Hamiltonian level before deriving the evolution equation
  of orbital elements
  leads to the erroneous conclusion that the projected angular momenta
  (i.e., $H_0^{\ast\ast}$ and $H_3^{\ast\ast}$) are conserved.
Alternatively, if the DA process is carried out on the equation of motion,
  this assumption is permissible prior to DA procedure,
  as typically done in the Lagrangian perturbation method \citep{Lim2020}
  or those solving Hamiltonian system
  via the vectorial form of equation of motion \citep[e.g.,][]{Liu2015}.
This assumption, which greatly simplifies calculations,
  is related to the choice of reference plane,
  specifically one that is perpendicular to
  the total angular momentum \cite{Murray2000,Naoz2013a},
  as illustrated in Fig.~\ref{fig:geometry}.
This reference plane is commonly referred to as the {\it invariable plane} \citep{Murray1999},
  offering benefits such as
  maintaining the relative inclination angle identical to the total inclination (i.e. $\iota \equiv \iota_0 + \iota_3$),
  and ensuring well-defined conservation of the total angular momentum.
Since canonical transformations preserve Poincaré invariance,
  this conservation of angular momentum
  persists for the DA orbital elements,
  implying that $\delta \Omega^{\ast\ast} = \pi$
  and $\delta \Omega = \pi$ are two interchangeable and equivalent conditions.
A more comprehensive discussion of subtle issues relevant
  is provided in Appendix~\ref{ap:angularMomentum}.

While the conservation of $H_0^{\ast\ast}$ (and $H_3^{\ast\ast}$)
  holds true in the test-particle-quadrupole limit \citep{Naoz2016},
  such conservation breaks down at octupole order when the outer binary
  is eccentric.
Beyond the test-particle-quadrupole limit,
  both $K^{\ast\ast}_{(i),n,m/2}$ and
  ${\partial K^{\ast\ast}}_{(i),n,m/2}/{\partial \delta \Omega^{\ast\ast}}$
  must be evaluated at $\delta \Omega^{\ast\ast} = \pi$
  to derive the correct evolution equation for the inclination angles.
\citet{Naoz2013a} posited that despite the incorrect treatment
  of this matter in the existing literature,
  it is still possible to recover the correct evolution equations
  by utilizing the conservation of total angular momentum.
Nevertheless, there are compelling reasons to defer
  the elimination of ascending nodes
  until the final stage of DA procedure.
Specifically, to obtain the correct cross term and generating function,
  the explicit formulas of
  ${\partial K}/{\partial \delta \Omega}$
  and ${\partial K^{\ast}}/{\partial \delta \Omega^{\ast}}$ are inevitable.
These cannot be obtained from
  existing literature where $\delta \Omega$ is not treated correctly.

Despite the great advantage of adopting $\delta \Omega = \pi$,
  this choice introduces potential issues
  when dissipative forces (e.g., gravitational radiation) are considered.
In the context of gravitational radiation,
  the inner binary generally radiates much more efficiently
  compared to the outer binary due to its compactness.
As a result, the total angular momentum is no longer conserved.
Enforcing $\delta \Omega = \pi$ is akin to
  describing the triple system relative to a reference plane that drifts
  in a complex and non-trivial manner.
This could introduce unphysical artefacts
  in gravitational wave waveform.
Similar artefacts caused by the dependence of multipoles
  on the centre-of-mass definition in triple systems,
  has been studied by \citep{Bonetti2017}.
Such limitations of existing DA methods
  need to be addressed to accurately model
  the orbital decay of a hierarchical triple system
  under gravitational radiation.
Potential solutions include deriving the Hamiltonian
  with explicit dependence on $\delta \Omega$,
  or continuously correcting the reference plane during integration.
We leave this for future studies.

Last but not least, the accurate construction
  of initial data proves crucial for
  the numerical stability of angular momentum conservation.
It is advantageous to calculate $\iota_3$ using
\begin{align}
 \iota_3 & = \sum_{n=1}^{\infty} \frac{(-1)^{n+1}}{n} \left( \frac{J_0}{J_3} \right)^n \sin n \iota \ ,
\end{align}
for precise initial data.
Such precise initial data is indispensable for evaluating the numerical stability
  and consequently, the reliability of the numerical solutions.
In our study, we used a $12(10)$th order
  explicit adaptive Runge-Kutta solver \cite{Feagin2012}
  for both the N-body simulation and secular perturbation theory
  to ensure robust conservation,
  although for the latter a lower-order solver suffices.
The maximum deviation of  $\delta \Omega$ were typically
  around $10^{-10} \, {\rm rad}$ and remained below $10^{-8} \,{\rm rad}$
  even in the worst-case scenario.
Additionally, energy and angular momentum were conserved at levels of $\sim 10^{-10}$ or better.

\subsection{Higher order terms} \label{sec:HigherOrderTerms}

\begin{table*}
\begin{ruledtabular}
\centering
\begin{tabular}{|c|c|c|c|c|}
\hline
 & \multicolumn{2}{c|}{$n=0$} & \multicolumn{2}{c|}{$n=1$} \\ \hline
 & $K^{\ast\ast}_{(0),n,m/2}/K^{\ast\ast}_{(0),0,0}$& $K^{\ast\ast}_{(i\ge1),n,m/2}/K^{\ast\ast}_{(0),0,0}$& $K^{\ast\ast}_{(0),n,m/2}/K^{\ast\ast}_{(0),0,0}$& $K^{\ast\ast}_{(i\ge1),n,m/2}/K^{\ast\ast}_{(0),0,0}$\\ \hline
$m=0$ & $1$ &  & $\delta$ &  \\ \hline
$m=1$ &  &  & $0$ &  \\ \hline
$m=2$ & $\alpha q_m$ &  & $\delta \alpha q_m$ &  \\ \hline
$m=3$ &  &  &  &  \\ \hline
$m=4$ &  &  & $\delta \alpha^2 {q_m}^2$ &  \\ \hline
$m=5$ &  &  & $\delta \alpha^{5/2} {q_m}^{3/2} (1-{e_3}^2)^{-1}$ &  \\ \hline
$m=6$ & $\alpha^3 q_m (1-{e_3}^2)^{-3/2}$ &  & $\delta \alpha^3 q_m (1-{e_3}^2)^{-3/2}$ & $\delta \alpha^3 q_m (1-{e_3}^2)^{-3} ({}^{\dagger})$\\ \hline
$m=7$ &  &  & $\delta \alpha^{7/2} {q_m}^{3/2} (1-{e_3}^2)^{-2} ({}^{\dagger}{}^{\ast}) $& $\delta \alpha^{7/2} {q_m}^{1/2} (1-{e_3}^2)^{-7/2} ({}^{\dagger})$\\ \hline
$m=8$ & $\alpha^4 q_m (1-{e_3}^2)^{-5/2} ({}^{\ast})$ &  & N/A & N/A \\ \hline
$m=9$ &  & $\alpha^{9/2} {q_m}^{3/2} (1-{e_3}^2)^{-3}$ & N/A & N/A \\ \hline
$m=10$ & $\alpha^5 q_m (1-{e_3}^2)^{-7/2}$ & $\alpha^5 q_m (1-{e_3}^2)^{-7/2} ({}^{\dagger})$& N/A & N/A \\ \hline
$m=11$ & & $\alpha^{11/2} {q_m}^{3/2} (1-{e_3}^2)^{-4} ({}^{\dagger}{}^{\ast})$& N/A & N/A \\ \hline
$m=12$ & $\alpha^6 q_m (1-{e_3}^2)^{-9/2}({}^{\ast})$ & $\alpha^6 {q_m}^2 (1-{e_3}^2)^{-9/2} ({}^{\dagger})$ & N/A & N/A \\ \hline
\end{tabular}
\caption{The dependency of direct and cross terms on the
  mass ratio and eccentricity is shown.
  For each order, only the dominant term
  for b-EMRI systems with $q_m \gg 1$ and $e_3 \sim 1$ is listed.
Terms marked with $({}^{\dagger})$ are estimates.
Terms marked with $({}^{\ast})$ depend on
  the mass difference of the inner binary (i.e., octupole-like).}
\label{Table:Order1}
\end{ruledtabular}
\end{table*}

The cross term due to interaction of
  $1$PN precession and quadrupole tidal force has gathered considerable
  attention, with several works \citep{Will2014a,Lim2020,Kuntz2023} dedicated to addressing this issue.
However, the complex nature of these cross terms makes it challenging
  to compare results across different studies and validate the findings,
  especially when comparing those of contact elements and osculating elements.
While our framework naturally facilitates the construction
  of such cross terms, we present only the leading-order periodic cross term at $(1,3/2)$ order,
  deferring the leading-order secular cross term, which is at $(1,3)$ order,
  for future studies.

Here, we provide a rough estimate of these higher-order cross terms
  to assess the necessity and potential for deriving them in b-EMRI type systems
  (i.e. $m_3 \gg m_2 \sim m_1 $).
Having established that higher-order cross terms contribute by either
  enhancing or suppressing the precession of the inner orbit,
  it is useful to estimate their contributions to $\dot{\omega}^{\ast\ast}_0$.
As evident from the construction of the canonical transformation, we have
\begin{align}
S_{n,\frac{m}{2}} \sim \frac{{a_0}^{3/2}}{M_0} K^{\ast\ast}_{n,\frac{m}{2}} \ , \nonumber \\
S^{\ast}_{n,\frac{m}{2}} \sim \frac{{a_3}^{3/2}}{M_3} K^{\ast\ast}_{n,\frac{m}{2}} \ .
\end{align}
Since the variation in the gravitational field due to the
  inner binary's periodic motion is smaller than that due to the outer binary,
  the cross terms generated from the second canonical transformation
  are generally more significant than those from the first canonical transformation.
Utilising Eq.~\ref{eq:SecondCanonicalK},
  we can estimate the contributions to precession at $(0,5)$ order to be
\begin{align}
\dot{\omega}_0^{\ast\ast} \big|_{(0,3)\hat{\otimes}(0,3)}  \sim
\sqrt{\frac{M_0}{{a_0}^3}}
\frac{\alpha^5 q_m}{(1-{e_3}^2)^{7/2}}   \ ,
\end{align}
which exhibits a similar dependence on eccentricity and mass ratio as
  the hexadecapole term.
Here, we use $\otimes$ to denote the cross term arising from the first
  canonical transformation, and $\hat{\otimes}$ for those originating from the
  second canonical transformation.
For $(0,11/2)$ order,
\begin{align}
\dot{\omega}_0^{\ast\ast} \big|_{(0,3)\hat{\otimes}(0,4)}  \sim
\sqrt{\frac{M_0}{{a_0}^3}}
\frac{\Delta m}{M_0}
\frac{\alpha^{11/2} {q_m}^{3/2}}{(1-{e_3}^2)^4} \ .
\end{align}
At the $(0,6)$ order, there are several contributions:
\begin{align}
& \dot{\omega}_0^{\ast\ast} \big|_{(0,9/2)\hat{\otimes}(0,3)}
  \sim
\dot{\omega}_0^{\ast\ast} \big|_{(0,3)\hat{\otimes}(0,3)\hat{\otimes}(0,3)}  \nonumber \\
 & \hspace*{2.35cm} \sim
\sqrt{\frac{M_0}{{a_0}^3}}
\frac{\alpha^{6} {q_m}^{2}}{(1-{e_3}^2)^{9/2}} \ ,
 \nonumber \\
& \dot{\omega}_0^{\ast\ast} \big|_{(0,3)\hat{\otimes}(0,4)} \sim
\sqrt{\frac{M_0}{{a_0}^3}}
\frac{\Delta m}{M_0}
\frac{\alpha^{6} q_m}{(1-{e_3}^2)^{9/2}} \ , \nonumber \\
& \dot{\omega}_0^{\ast\ast} \big|_{(0,3)\otimes(0,3)}  \sim
\sqrt{\frac{M_0}{{a_0}^3}}
\frac{\alpha^{6} {q_m}^{2}}{(1-{e_3}^2)^{3}} \ .
\end{align}
The dominate contribution comes from the
  second order cross term (e.g., three-term cross term).

There are several interaction terms contributing at $(1,3)$ order,
  among which the dominate contribution come from the three-term cross term, and it is given by
\begin{align}
\label{eq:triple}
\dot{\omega}_0^{\ast\ast} \big|_{(0,3)\hat{\otimes}(0,3)\hat{\otimes}(1,0)}  \sim
   \sqrt{\frac{M_0}{{a_0}^3}} \frac{\delta \alpha^{3} q_m}{(1-{e_3}^2)^3} \ .
\end{align}
There are also $(1,3)$ order cross term
  resulting from $(1,0)\otimes(0,3)$ and $(1,1)\hat{\otimes}(0,3)$.
Both of these have
\begin{align}
\dot{\omega}_0^{\ast\ast} \big|_{(1,0)\otimes(0,3)}
 & \sim
\dot{\omega}_0^{\ast\ast} \big|_{(1,1)\hat{\otimes}(0,3)}
\nonumber \\
& \sim
   \sqrt{\frac{M_0}{{a_0}^3}} \frac{\delta \alpha^{3} q_m}{(1-{e_3}^2)^{3/2}} \ .
\end{align}
Such estimation agrees with the calculation of \citep{Will2014a}.
We further estimated the direct and cross terms at $(1,7/2)$ order
  and list out all the findings in table.~\ref{Table:Order1}.

Comparing Eq.~\ref{eq:triple} with Eq.~\ref{eq:omegaDot} suggests
  that for the $(1,3)$ order to have a noticeable effect on the
  precession, we need to have a sufficiently large $\delta (1-{e_3}^2)^{-3/2}$,
Hence, the $(1,3)$ order cross term is likely to become interesting
  when $e_3 \to 1$ and $\delta \to 1$.
While we observe that the DA method can
  remain valid even for highly eccentric third bodies (e.g., $e_3 = 0.89$ in Fig.~\ref{fig:CrossDirectCompare}),
  when the eccentricity is more extreme, the angular velocity of
  the outer binary would be comparable to the inner binary,
  indicating that the DA method would likely break down
  (see Sec~\ref{sec:Limitation} for details).
Since the $(1,3)$ order arises from the periodic part of
  the quadrupole and 1PN perturbation, it is connected to the deviation
  of the perturbed orbit from a Keplerian ellipse.
In the case of 1PN perturbation, significant deviation occurs only when the associated precession
  is substantial.
This implies that if this effect is strong enough to modify the evolution
  of the triple system, the relativistic precession likely has destroyed the \ac{vzlk} resonance.
The contradictory conditions necessary for the $(1,3)$ order to be significant
  suggest that this cross term is unlikely to have a
  noticeable effect on the evolution of the triple system
  in the classical \ac{vzlk} domain or a highly-relativistic system.
However, this does not rule out the possibility of the $(1,3)$ order
  having a significant impact near its resonances in weakly-relativistic systems.

These resonances are particularly feasible when
  relativistic precession counteracts and suppresses
  the quadrupole oscillation, extending the quadrupole time scale.
This extended time scale can potentially lead to
  the excitation of resonances involving higher-order terms
  that would otherwise be dampened by the quadrupole term.
Moreover, \citep{Kuntz2023} pointed out that
   this cross term could retrigger orbital flip that are otherwise
   suppressed by relativistic interaction.
Investigating this issue is beyond the scope of this work,
  and we eagerly anticipate future research exploring this topic.

\subsection{Limitation of secular perturbation theory:
  mean motion resonance, energy exchange and stability}
\label{sec:Limitation}

The system evolution
  predicted from secular perturbation theory
  diverges from those of N-body simulations when approaching
   the unstable parameter space
   \citep[see e.g.][]{Antonini2012}.
This region is typically characterised by
  the presence of a closer, more eccentric, and more massive third body,
  in which case the system exhibits chaotic energy exchange
  and large eccentricity jumps during the energy exchange.
As we shall see in this section,
  such break down of secular perturbation theory
  is related to the \ac{mmr}s,
  which is related to many subtle issues with secular perturbation theory framework,
  including the difference in time scale of \ac{vzlk} oscillation when compared
  against N-body simulation.

The concept of \ac{mmr} is mentioned very often
  in the context of planetary systems, where planets are found
  in small, e.g., $2/1$ or $3/2$
 resonances with each other,
  where the orbits are not particularly hierarchical.
In the context we are interested in, the situation is different,
  but both of these cases can be well explained by the
  same theory, using the simple model of the forced non-linear harmonic oscillator
  and the overlap of resonance \citep[see lecture notes][for details]{Mardling2008},
  which we briefly recap here.

\begin{figure}
\vspace*{-1cm} \centering
\includegraphics[width=1\columnwidth]{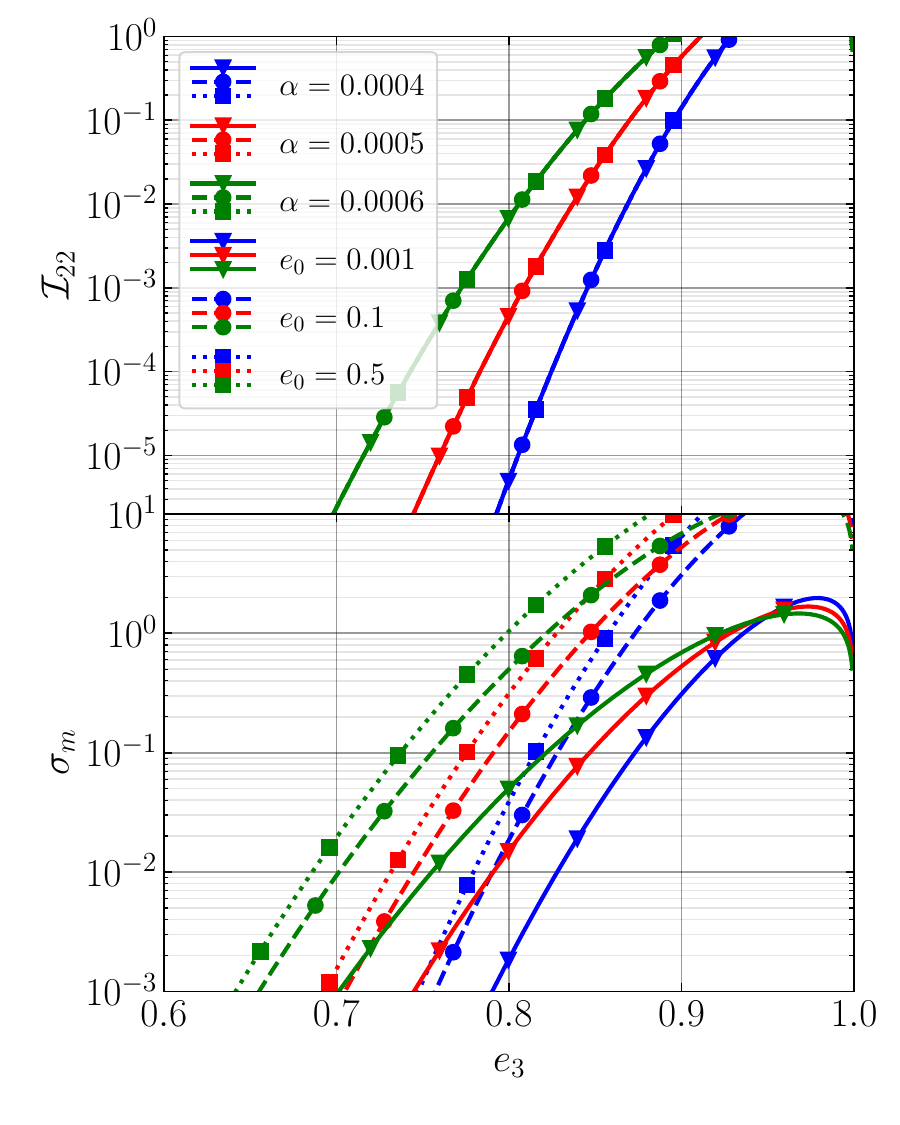}
\vspace*{-1cm}
  \caption{The top panel shows the parameter for energy exchange
    and eccentricity oscillation during periapsis passage
    of the third body.
The bottom panel shows the width of the resonance.
As chaotic energy exchanges emerge for $\sigma_m \ge 1/2$,
  systems with larger eccentricities (either initial or due to excitation)
  and less hierarchical structures (i.e., larger $\alpha$)
  are more likely to experience such exchanges.}
  \label{fig:resonanceOverlap}
\end{figure}

While a canonical transformation, in principle, should keep all
  the features, including resonances and symmetry intact,
  the separation of time scale in the canonical transformation,
  despite being a very efficient and elegant
  way of averaging the system evolution and sorting the strength
  of perturbation, breaks and eliminates several resonances.
The two sequential canonical transformations
  separate the motion of the inner binary from the outer binary,
  and further from any other time scale within the system,
  effectively decoupling the two systems and leading to an adiabatic secular evolution.
For example, the resonance between the outer binary's mean motion
  and the relativistic precession of the inner binary \citep[see, e.g.,][for the treatment of this resonance]{Kuntz2022} is ignored in the current secular perturbation framework.
This is true for most of the other orbital elements as well --
  while the periodic variation of these orbital elements induced by
  a varying gravitational field has been taken care of by the cross terms,
  the secular variation of them has been eliminated by
  the canonical transformation process, during which the integration is performed
  assuming others are constant.
While these resonances may not appear concerning as their secular evolution
  time scale appears much longer than the orbital time scale,
  and hence the corresponding resonance is unlikely to be important,
  the \ac{mmr} could be important for b-EMRI system with a very eccentric third body.
The elimination of the \ac{mmr} can be seen from the simple argument below:
if we write the quadrupole Hamiltonian in the Fourier series,
  the SA Hamiltonian is simply the zeroth order,
  and the generating function is simply
\begin{align}
K_{(0),0,3}
  &  = \sum_{n,m} F_{n,m} e^{i n l_0} e^{i m l_3} \ ,  \nonumber \\
K^{\ast}_{(0),0,3}
  & = \sum_{m} F_{0,m} e^{i m l_3} \ ,  \nonumber \\
S_{(0),0,3}
  &  = \sum_{n\neq 0}
    \frac{i F_{n,m}}{n n_0} e^{i n l_0} e^{i m l_3}  \ ,
\end{align}
where $n_0 \equiv 2\pi/P_0$ represents the mean motion,
  and $F_{n,m}$ are functions of the other orbital elements.
However, if we include the motion of the outer binary,
  the generating function should take the form
\begin{align}
\mathcal{S}_{(0),0,3}
    = \sum_{n,m,nm\neq 0} \frac{i F_{n,m}}{n n_0 + m n_3}
      e^{i n l_0} e^{i m l_3}  \ .
\end{align}
As $n_3 \ll n_0$, $S_{(0),0,3}$ is simply the zeroth
  order of the Taylor series of $\mathcal{S}_{(0),0,3}$
  when expanded with respect to $n_3/n_0$,
  and the remaining part of the series are smeared out
  in higher order terms.
Particularly, the next order is hidden in $S_{(1),0,9/2}$,
 where
\begin{align}
S_{(1),0,9/2} & = -\frac{K'_{0,2}(L_3)}{K'_{0,0}(L_0)}
 \bigg\{ \frac{\partial S_{(0),0,3}}{\partial l_3} \bigg\}_{l_0}
  \nonumber \\
& = \sum_{n,m,nm\neq 0} \frac{i m n_3 F_{n,m}}{n^2 {n_0}^2}
       e^{i n l_0} e^{i m l_3}  \ .
\end{align}
In fact, it is by enforcing this series expansion
  that the \ac{mmr} is eliminated.
We note that there always exists a combination
  $(n:m)= (\pm 1:m)$
  which allows for $|\pm n_0 + m n_3| < n_3$ that do not get eliminated
  on the orbital time scale of the outer binary,
  which is precisely the \ac{mmr} mentioned above.
In the mild eccentricity $e_3$ regime, the Fourier coefficient
  decays rapidly towards large $|m|$ and the resonance
  can only be excited when $n_0/n_3$ is relatively small.
This regime usually appears in the planetary system.
However, in our system where $e_3 \sim 1$,
  the Fourier coefficients decay slowly towards large $m$,
  and such that even very hierarchical system exhibit \ac{mmr}s.
Further, the high eccentricity complicates the calculation
  of such resonances due to the slow convergence of the Fourier
  series, rendering the computation different from that of
  planetary system.
In fact, this situation resembles more of a flyby event.
A good approximation of the eccentricity oscillation
  and energy exchange during such a close encounter
  can be found in \citep{Mardling2008}, where
\begin{align}
& \frac{\Delta E_0}{E_0}
     \simeq {\mathcal{I}_{22}}^2 + 2 e_0 \mathcal{I}_{22} \sin \phi_{2n1} \ ,  \nonumber \\
& e_0(P_3)  \simeq
   \sqrt{e_0(0)^2 - 2 e_0(0) \mathcal{I}_{22} \sin \phi_{2n1}
   + {\mathcal{I}_{22}}^2 } \ ,
\end{align}
 with
\begin{align}
& \phi_{2n1}  = l_0(0) + \sigma \pi + 2 (\omega_0 - \omega_3)\ ,
  \nonumber \\
& \sigma  = n_0 /n_3 \ ,  \nonumber \\
&\mathcal{I}_{22}  = \frac{9}{4} \frac{m_3}{M_0}
   \left( \frac{a_0}{a_3} \right)^3 \mathcal{E}_{22} \ ,
    \nonumber \\
& \mathcal{E}_{22}  = n_0 e^{-i \sigma \pi} \int_0^{P_3} {\rm d}t \
    \frac{e^{-2i \phi_3}}{(R_3/a_3)^3}\  e^{i n_0 t}   \ .
\end{align}
The asymptotic form of  the integral $\mathcal{E}_{22}$ is
\begin{align}
\mathcal{E}_{22}  \simeq \frac{4\sqrt{2\pi}}{3} \frac{(1-{e_3}^2)^{3/4}}{{e_3}^2}
         \sigma^{5/2} e^{-\sigma \xi(e_3)} \ ,
\end{align}
where $\xi(e_3)  \equiv \cosh^{-1}(1/e_3) - \sqrt{1-{e_3}^2}$.
The exponent $\sigma \xi(e_3)$ dictates
  the strength of the energy exchange.
We notice that in the $e_3 \to 1$ limit one finds
\begin{align}
\lim_{e_3 \to 1} \sigma \xi(e_3) = \frac{1}{3} \frac{P_3}{P_0} (1-{e_3}^2)^{3/2} \simeq \frac{1}{3 \Upsilon} \,,
\end{align}
(see Eq.~\eqref{eq:Ysilon} for the expression of $\Upsilon$),
indicating clearly the relationship between
  the strength of the tidal force and the energy exchange
  during the close encounter.
The energy exchange could be periodic or chaotic,
  with the boundary described by the overlapping of nearby
  resonance, which can roughly be described as $\sigma_m \sim 1/2$,
  with $\sigma_m$ being the width of the $(1:m)$ resonance:
\begin{align}
&\sigma_m  = 2 \sqrt{
\frac{9}{2} s_1^{(22)}(e_0) F_m^{(22)}(e_3) \left[ \frac{m_3}{M_3}
+ \frac{\mu_0}{M_0} \left(\frac{m\,M_0}{M_3}\right)^{2/3} \right]} \ , \nonumber \\
&s_1^{(22)}(e_0)  \simeq 3 e_0 - \frac{13}{8} {e_0}^3
    - \frac{5}{192} {e_0}^5 + \frac{227}{3072} {e_0}^7 \ ,
     \nonumber \\
&F_m^{(22)}(e_3)  \simeq \frac{\mathcal{E}_{22} (e_3, m)}{2 \pi m } \ .
\end{align}
For the system parameters shown in Fig.~\ref{fig:KastastK3Newtonian},
  the values of $\mathcal{I}_{22}$ and $\sigma_m$
  are shown in Fig.~\ref{fig:resonanceOverlap}.
For the system with small eccentricities $e_0 \sim 0.001$,
  the system appears to have $\sigma_m < 0.5$
  for $e_3 = 0.8$,
  suggesting that in this domain the system's energy exchange
  and eccentricity oscillations due to the \ac{mmr}
  appear to be (semi-)periodic.
When $e_0 \sim 0.1$, the system with $\alpha = 0.0006$
  appears to be in the resonance overlap regime,
  whereas the systems with $\alpha = 0.0005$ and $\alpha = 0.0004$
  appear to exhibit (semi-)periodicity.

\section{Summary} \label{sec:summary}

In this work we generalized von Zeipel's secular perturbation theory
  and constructed a Hamiltonian framework for the secular evolution of
  relativistic hierarchical triple systems, with a special focus on b-EMRI systems.
We utilized the properties of canonical transformations to derive
  the generating functions and \acl{da} Hamiltonian via a two-factor series expansion.
During the canonical transformations, by eliminating the fast-oscillating mean anomalies,
  the high-frequency oscillations of the orbital elements over orbital time scales are averaged out,
  leaving only the slow evolution.
This process clarifies a subtlety in secular perturbation theory:
  it predicts the evolution of the averaged orbital elements rather than
  the true orbital elements.
This correspondence is particularly important for comparing predictions
  from secular perturbation theory with those from N-body simulations.
As for the \ac{da} Hamiltonian, we
  recover the secular direct terms up to the $(0,4)$ and $(1,2)$ order reported in \cite{Naoz2013b}
  the quadrupole-squared cross term at order $(0,9/2)$ reported by \citep{Will2021}.
Additionally, we identify a new direct term at $(1,5/2)$ order related to spin-orbit coupling
  and two periodic cross terms at $(1,3/2)$ and $(1,5/2)$ orders.
While these periodic cross terms do not directly affect secular evolution,
  they are crucial for the correspondence between the orbital elements
  and the averaged orbital elements.

With these new direct and cross terms, we perform several case studies to
  explore their physical implications and
  identify resonances caused by their interactions.
By comparing our results to N-body simulations,
  we confirmed the presence of resonance arising from
  the interplay between the quadrupole-squared cross term and the quadrupole term.
The discrepancies observed between the N-body simulations
  and secular perturbation theory suggest that overlooked ingredients,
  such as mean-motion resonances (MMRs),
  are significant in parameter spaces where the quadrupole-squared term is important.
This underscores the challenges of developing
  a precise secular perturbation theory
  that can fully predict the evolution of a triple system.
The chaotic nature of the system dynamics adds further complexity,
 urging us to find appropriate methods to compare secular perturbation theory with N-body
 simulations and reassess the capabilities and limitations of secular perturbation theory.
The complex interplay of relativistic terms with Newtonian terms
 indicates the difficulty of dealing with relativistic triple systems,
 as resonances become ubiquitous and can emerge from the interaction of
 multiple interaction terms.

Additionally, we briefly commented on the
  differences between Hamiltonian and Lagrangian approaches to
  secular perturbation theory.
We also discussed the technical and numerical subtleties in eliminating the ascending nodes
  and conserving angular momentum.
We estimated the magnitude of direct and cross terms of orders above $(0,9/2)$ and $(1,5/2)$.
Furthermore, we examined the fundamental limitations of secular perturbation theory,
  highlighting the fact that efforts to include \ac{mmr}s must be undertaken in parallel with efforts to
  extend the theory to include higher-order terms
  like the quadrupole-cubic term.
One possibility is to
  refine the generating functions in order to account for
  the motion of the outer binary
  \citep[see, e.g.,][]{Lei2019}.
  Another approach is to incorporate the secular effects of energy and angular momentum exchange
  during close flybys \citep[e.g.,][]{Mardling2008}.
 Further work on this topic will be published in future studies.

\begin{acknowledgments}
We would like to express our gratitude for
  the insightful discussions with participants of
  the ``Theories of the three-body system'' workshop
  held at the Monash Prato Centre.
In particular, we thank Rosemary Mardling
  for valuable discussions and
  for her insights into this work.
We thank Paul Lai for proofreading this manucript.
KJL appreciates
  the hospitality of the Institute of Theoretical Physics of
  Chinese Academic of Science,
  which facilitated the completion of part of this research.
KJL was supported by a PhD Scholarship from the Vinson and Cissy Chu Foundation, and by a University College London (UCL)
MAPS Dean’s Prize.
KW acknowledges support from the UCL Cosmoparticle Initiative.
ZY was supported by a United Kingdom Research and Innovation (UKRI) Stephen Hawking Fellowship.
This work utilized {\it Mathematica}, {\it Python},
  and several packages, including {\it matplotlib} and {\it numpy},
  as well as the Mathematica package {\it Format} \citep{Sofroniou1994}.
  This work has made use of NASA's Astrophysics Data System (ADS).
\end{acknowledgments}

\onecolumngrid
\appendix

\section{Conservation of angular momentum}
\label{ap:angularMomentum}

While canonical transformations preserve the Poincar\'e
  symmetry, including the rotational symmetry which is manifested by angular momentum
  conservation, subtle issue arises due to the truncation the of Taylor series at finite order.
The truncated DA Hamiltonian, being an approximation
  itself, suggests that energy is only approximately conserved,
  and further, the angular momentum conservation also becomes an approximation.
Therefore, it would be useful to assess to which degree is
  the angular momentum conserved for the cross term presented above.
Due to the symmetry of the problem, we choose to assess
  this for the second canonical transformation,
  and comment that conclusion remain intact
  for the first canonical transformation.
Firstly, we introduce a convenient operator
\begin{align}
  \nabla_{\boldsymbol{Q^{\ast}},\boldsymbol{P^{\ast\ast}}}  \equiv
  \frac{1}{2} \bigg\{ J^{\ast\ast}_0{}^2 \sin^2 \iota^{\ast\ast}_0
    - J^{\ast\ast}_3{}^2 \sin^2 \iota^{\ast\ast}_3 , \;\cdot\; \bigg\}_{\boldsymbol{Q^{\ast}},\boldsymbol{P^{\ast\ast}}}
  =  - J^{\ast\ast}_0 \frac{\partial}{\partial \omega^{\ast}_0} + H^{\ast\ast}_0
       \frac{\partial}{\partial \Omega^{\ast}_0}
  + J^{\ast\ast}_3 \frac{\partial}{\partial \omega^{\ast}_3} - H^{\ast\ast}_3
       \frac{\partial}{\partial \Omega^{\ast}_3} \ ,
\end{align}
which generalize the Poisson bracket to a
  mixture of old coordinate and new momenta.
We reserve the symbol $\nabla_{\boldsymbol{Q^{\ast\ast}},\boldsymbol{P^{\ast\ast}}}$
   as a shorthand for
  the corresponding Poisson bracket with respect to basis $(\boldsymbol{Q^{\ast\ast}},\boldsymbol{P^{\ast\ast}})$,
  and similarly for $\nabla_{\boldsymbol{Q^{\ast}},\boldsymbol{P^{\ast}}}$.

For an arbitrary function $A(\boldsymbol{Q^{\ast}},\boldsymbol{P^{\ast\ast}})$,
  this operator satisfy that
\begin{align}
& \nabla_{\boldsymbol{Q^{\ast}},\boldsymbol{P^{\ast\ast}}}  \big\{ A \big\}_{l^{\ast}_3}
   = \big\{ \nabla_{\boldsymbol{Q^{\ast}},\boldsymbol{P^{\ast\ast}}} A \big\}_{l^{\ast}_3} \ , \nonumber \\
& \nabla_{\boldsymbol{Q^{\ast}},\boldsymbol{P^{\ast\ast}}}
   \frac{\partial A}{\partial \boldsymbol{P^{\ast\ast}}}
   =  - \frac{\partial A}{\partial \boldsymbol{Q^{\ast}}}
      \nabla_{\boldsymbol{Q^{\ast}},\boldsymbol{P^{\ast\ast}}}
      \bigg( \sum_i \frac{\boldsymbol{Q^{\ast}}_i}{\boldsymbol{P^{\ast\ast}}_i} \bigg) \ , \nonumber \\
& \nabla_{\boldsymbol{Q^{\ast}},\boldsymbol{P^{\ast\ast}}}
    \frac{\partial A}{\partial \boldsymbol{Q^{\ast}}}
   = \frac{\partial }{\partial \boldsymbol{Q^{\ast}}} \nabla_{\boldsymbol{Q^{\ast}},\boldsymbol{P^{\ast\ast}}}A
     \ .
\end{align}
The conservation of angular momentum during the canonical transformation
  and the conservation of total angular momentum
  are intricately interconnected,
  representing different facets of a unified concept.
The angular momentum conservation during the second canonical transformation
  can be written as
\begin{align}
& J^{\ast}_0{}^2  \sin^2 \iota^{\ast}_0   - J^{\ast}_3 {}^2
   \sin^2 \iota^{\ast}_3
 =  J^{\ast\ast}_0{}^2 \sin^2 \iota^{\ast\ast}_0   - J^{\ast\ast}_3{}^2 \sin^2 \iota^{\ast\ast}_3  +
\sum_{n,m,i} \Pi^{\ast}_{(i),n,\frac{m}{2}} (\boldsymbol{Q^{\ast}},\boldsymbol{P^{\ast\ast}}) \, {\rm with}
  \nonumber \\
& \Pi^{\ast}_{(0),n,\frac{m}{2}}  =
- 2 \nabla_{\boldsymbol{Q^{\ast}},\boldsymbol{P^{\ast\ast}}} S^{\ast}_{(0),n,\frac{m}{2}} \ ,\nonumber  \\
&  \Pi^{\ast}_{(1),n,\frac{m}{2}}  =
- 2 \nabla_{\boldsymbol{Q^{\ast}},\boldsymbol{P^{\ast\ast}}} S^{\ast}_{(1),n,\frac{m}{2}}
+ \sum_{k,l} \bigg(
   \frac{\partial S^{\ast}_{(0),k,\frac{l}{2}}}{\partial \omega^{\ast}_0}
   \frac{\partial S^{\ast}_{(0),n-k,\frac{m-l+3}{2}}}{\partial \omega^{\ast}_0}
 - \frac{\partial S^{\ast}_{(0),k,\frac{l}{2}}}{\partial \Omega^{\ast}_0}
   \frac{\partial S^{\ast}_{(0),n-k,\frac{m-l+3}{2}}}{\partial \Omega^{\ast}_0} \nonumber \\
& \hspace*{5cm}
- \frac{\partial S^{\ast}_{(0),k,\frac{l}{2}}}{\partial \omega^{\ast}_3}
   \frac{\partial S^{\ast}_{(0),n-k,\frac{m-l+3}{2}}}{\partial \omega^{\ast}_3}
 + \frac{\partial S^{\ast}_{(0),k,\frac{l}{2}}}{\partial \Omega^{\ast}_3}
   \frac{\partial S^{\ast}_{(0),n-k,\frac{m-l+3}{2}}}{\partial \Omega^{\ast}_3}
\bigg) \ .
\end{align}
For the zero-th order, we have
\begin{align}
& \nabla_{\boldsymbol{Q^{\ast}},\boldsymbol{P^{\ast\ast}}}
    K^{\ast\ast}_{(0),n,\frac{m}{2}} 
     = \left< \nabla_{\boldsymbol{Q^{\ast}},\boldsymbol{P^{\ast\ast}}}
       K^{\ast}_{n,\frac{m}{2}} 
         \right>_{l^{\ast}_3} = 0   \ ,  \nonumber \\
& \nabla_{\boldsymbol{Q^{\ast}},\boldsymbol{P^{\ast\ast}}}
    S^{\ast}_{(0),n,\frac{m}{2}}
     = \frac{1}{K^{\ast\prime}_{0,1}(L^{\ast}_3)} \bigg\{ \int {\rm d}
      l^{\ast}_3
\big\{ \nabla_{\boldsymbol{Q^{\ast}},\boldsymbol{P^{\ast\ast}}} K^{\ast}_{n,\frac{m}{2}}
   \big\}_{l^{\ast}_3}  \bigg\}_{l^{\ast}_3}  = 0 \ .
\end{align}
Whereas for the first-order,
  it can be easily shown that, our choice of generating function
  in Eq.~\ref{eq:secondDirect} perfectly cancels out
  the other component inside
  $\Pi^{\ast}_{(1),n,m/2}$, causing it to vanish completely.

Similarly, for the the first order Hamiltonian,
  our choice of generating function satisfies
\begin{align}
 \nabla_{\boldsymbol{Q^{\ast}},\boldsymbol{P^{\ast\ast}}}
  C^{\ast}_{(1),n,\frac{m}{2}}
& =  \nabla_{\boldsymbol{Q^{\ast}},\boldsymbol{P^{\ast\ast}}}
  \sum_{k,l,(k,l)\neq(0,2)} \bigg(
  \frac{\partial C^{\ast}_{(0),k,\frac{l}{2}}}{\partial \boldsymbol{P^{\ast\ast}}_0}
          \frac{\partial S^{\ast}_{(0),n-k,\frac{m-l}{2}}}{\partial \boldsymbol{Q^{\ast}}_0}
- \frac{\partial K^{\ast\ast}_{(0),k,\frac{l}{2}}}{\partial \boldsymbol{Q^{\ast}}_0}
          \frac{\partial S^{\ast}_{(0),n-k,\frac{m-l}{2}}}{\partial \boldsymbol{P^{\ast\ast}}_0}
         \nonumber  \\
& \hspace{2cm}
 + \frac{\partial C^{\ast}_{(0),k,\frac{l}{2}}}{\partial \boldsymbol{P^{\ast\ast}}_3}
          \frac{\partial S^{\ast}_{(0),n-k,\frac{m-l-1}{2}}}{\partial \boldsymbol{Q^{\ast}}_3}
 - \frac{\partial K^{\ast\ast}_{(0),k,\frac{l}{2}}}{\partial \boldsymbol{Q^{\ast}}_3}
          \frac{\partial S^{\ast}_{(0),n-k,\frac{m-l-1}{2}}}{\partial \boldsymbol{P^{\ast\ast}}_3}
           \nonumber \\
& \hspace*{2cm}
+ \frac{1}{2} \frac{\partial^2 K^{\ast}_{0,1}}{\partial L^{\ast\ast}_3{}^2}
          \frac{\partial S^{\ast}_{(0),k,\frac{l}{2}}}{\partial l^{\ast}_3} \frac{\partial S^{\ast}_{(0),n-k-i,\frac{m-l}{2}-2}}{\partial l^{\ast}_3}
 \bigg)   \nonumber \\
 &  = - \frac{1}{2} \frac{\partial K^{\ast}_{0,1}}{\partial L^{\ast\ast}_3}
    \sum_{k,l,(k,l)\neq (0,2)}
  \frac{\partial }{\partial l^{\ast}_3}
  \bigg(\frac{\partial S^{\ast}_{k,\frac{l}{2}}}{\partial \omega^{\ast}_0} \frac{\partial S^{\ast}_{n-k,\frac{m-l}{2}}}{\partial \omega^{\ast}_0}
 - \frac{\partial S^{\ast}_{k,\frac{l}{2}}}{\partial \Omega^{\ast}_0} \frac{\partial S^{\ast}_{n-k,\frac{m-l}{2}}}{\partial \Omega^{\ast}_0}
\nonumber \\
& \hspace*{2cm}
  - \frac{\partial S^{\ast}_{k,\frac{l}{2}}}{\partial \omega^{\ast}_3}
     \frac{\partial S^{\ast}_{n-k,\frac{m-l-1}{2}}}{\partial \omega^{\ast}_3}
+ \frac{\partial S^{\ast}_{k,\frac{l}{2}}}{\partial \Omega^{\ast}_3} \frac{\partial S^{\ast}_{n-k,\frac{m-l}{2}}}{\partial \Omega^{\ast}_3}
  \bigg) \nonumber  \\
&
  = - \frac{\partial K^{\ast}_{0,1}}{\partial L^{\ast\ast}_3}
  \frac{\partial }{\partial l^{\ast}_3}  \bigg( \nabla_{\boldsymbol{Q^{\ast}},\boldsymbol{P^{\ast\ast}}}
 S^{\ast}_{(1),n,\frac{m}{2}} \bigg)
\end{align}
 such that
\begin{align}
\nabla_{\boldsymbol{Q^{\ast}},\boldsymbol{P^{\ast\ast}}}
   K^{\ast\ast}_{(1),n,\frac{m}{2}}  =
   \left\langle \nabla_{\boldsymbol{Q^{\ast}},\boldsymbol{P^{\ast\ast}}}
  C^{\ast}_{(1),n,\frac{m}{2}} \right\rangle_{l^{\ast}_3}
  = 0    \ .
\end{align}
Similar proof exists for the first non-trivial canonical transformation.
Further, for the second order Hamiltonian, we have
\begin{align}
\nabla_{\boldsymbol{Q^{\ast}},\boldsymbol{P^{\ast\ast}}}
  C^{\ast}_{(2),n,\frac{m}{2}}
  = \eth (\omega^{\ast}_0) - \eth (\Omega^{\ast}_0) - \eth (\omega^{\ast}_3) + \eth (\Omega^{\ast}_3)
\end{align}
 where
\begin{align}
\eth (\omega^{\ast}_0) &  \equiv
  \sum_{i,j,k,l,(i,j)\neq (0,2)}
  \Bigg[
  \frac{\partial^2 K^{\ast}_{(0),i,\frac{j}{2}}}{\partial \omega^{\ast}_0
                \partial \boldsymbol{P^{\ast\ast}}}
    \frac{\partial S^{\ast}_{(1),k,\frac{l}{2}}}{\partial \omega^{\ast}_0}
    \frac{\partial S^{\ast}_{(1),q,\frac{p}{2}}}{\partial \boldsymbol{Q^{\ast}}}
- \frac{\partial^2 K^{\ast\ast}_{(0),i,\frac{j}{2}}}{\partial \omega^{\ast}_0
                  \partial \boldsymbol{Q^{\ast}}}
    \frac{\partial S^{\ast}_{(1),k,\frac{l}{2}}}{\partial \omega^{\ast}_0}
    \frac{\partial S^{\ast}_{(1),q,\frac{p}{2}}}{\partial \boldsymbol{P^{\ast\ast}}} \Bigg]
     \nonumber \\
    & \hspace*{0.5cm}
    + \sum_{k,l,(k,l)\neq (0,2)} \bigg[
\frac{\partial K^{\ast}_{(0),k,\frac{l}{2}}}{\partial \omega^{\ast}_0}
    \frac{\partial S^{\ast}_{(1),q,\frac{p}{2}}}{\partial \omega^{\ast}_0}
+ \frac{1}{2} \frac{\partial K^{\ast}_{(0),k,\frac{l}{2}}}{\partial \boldsymbol{P^{\ast\ast}}}
    \frac{\partial }{\partial \boldsymbol{Q^{\ast}}} \bigg( \frac{\partial S^{\ast}_{(0),i,\frac{j}{2}}}{\partial \omega^{\ast}_0} \frac{\partial S^{\ast}_{(0),q,\frac{p}{2}}}{\partial \omega^{\ast}_0} \bigg)
    \nonumber \\
& \hspace*{1.5cm}
 - \frac{\partial K^{\ast\ast}_{(0),k,\frac{l}{2}}}{\partial \omega^{\ast}_0}
    \frac{\partial S^{\ast}_{(1),q,\frac{p}{2}}}{\partial \omega^{\ast}_0}
- \frac{1}{2} \frac{\partial K^{\ast\ast}_{(0),k,\frac{l}{2}}}{\partial \boldsymbol{Q^{\ast}}}
    \frac{\partial }{\partial \boldsymbol{P^{\ast\ast}}} \bigg( \frac{\partial S^{\ast}_{(0),i,\frac{j}{2}}}{\partial \omega^{\ast}_0} \frac{\partial S^{\ast}_{(0),q,\frac{p}{2}}}{\partial \omega^{\ast}_0} \bigg) \nonumber \\
& \hspace*{1.5cm}
  - \frac{\partial S^{\ast}_{(0),q,\frac{p}{2}}}{\partial \omega^{\ast}_0}
     \frac{\partial }{\partial \omega^{\ast}_0} 
    \bigg( C^{\ast}_{(1),k,\frac{l}{2}} +
     \frac{\partial K^{\ast}_{0,1}}{\partial L^{\ast\ast}_3}
         \frac{\partial S^{\ast}_{(1),k,\frac{l}{2}}}{\partial l^{\ast}_3} \bigg)
  \bigg]  \ .
\end{align}
Here $q$ and $p$ should be the appropriate integers used to match the right order,
  that may differ from term by term.
It is straight forward to verify that our choice of Eq.~\ref{eq:secondDirect} is indeed
  one solution that satisfy
\begin{align}
  \left\langle \nabla_{\boldsymbol{Q^{\ast}},\boldsymbol{P^{\ast\ast}}}
  C^{\ast}_{(2),n,\frac{m}{2}} \right\rangle_{l^{\ast}_3}
  = 0 \ .
\end{align}
Consequently, we demonstrate that
  by adjusting the generating function to permit a constant offset
  between the contact elements and DA orbital elements, we guarantee
  the conservation of angular momentum during the canonical transformation
  at the first order, and the conservation of angular momentum within the Hamiltonian at the second order.
This conservation ensures that no artificial rotation
  of the reference plane is introduced during the canonical transformation,
  allowing us to simplify the Hamiltonian by eliminating the two ascending nodes.

\section{Formula} \label{ap:formula}

Here we explicitly list the secular interaction terms and relativistic binary terms
  up to $(0,9/2)$ order for Newtonian interactions,
  and up to $(1,5/2)$ order for relativistic interactions.
Our Hamiltonian has been verified up to $(0,3)$ order and $(1,2)$ order,
  in agreement with \citep{Naoz2013b},
  and our secular orbital equation has been validated against \citep{Will2021}.
To the best of our knowledge,
  currently, no existing formula for
  $({\partial K^{\ast\ast}}/{\partial \delta \Omega})$
  has been reported in the literature.
It is important to note that the orbital elements
  presented here are the DA ones, with the upper indices omitted for brevity.
They are
\begin{align}
& K^{\ast\ast}_{({\rm b}),1,0}=\frac{\mu_0 M_0 }{8 {a_0}^2 J_0} \big[3 M_0 \big(5 J_0-8 L_0\big)-J_0 \mu_0\big]
   \ ,  \text{and $(0 \to 3)$ for $K_{({\rm b}),1,2}^{\ast,\ast}$} \ , \nonumber \\
& \frac{\partial K^{\ast\ast}_{({\rm b}),1,0}}{\partial \delta \Omega}=0\ ,  \nonumber \\
& K^{\ast\ast}_{(0),0,3}=\frac{{\mu_3}^6 {L_0}^2 m_3 {M_3}^3 }{32 {J_3}^3 {\mu_0}^3 {L_3}^3 {M_0}^2} \big[\big(3 \cos 2\iota+1\big) \big(3 {J_0}^2-5 {L_0}^2\big)-30 a_0 {\mu_0}^2 M_0 {e_0}^2 \sin^2 \iota \cos 2\omega_0\big]\ ,
  \nonumber \\
& \frac{\partial K^{\ast\ast}_{(0),0,3}}{\partial \delta \Omega}=\frac{15 a_0 {\mu_3}^6 {L_0}^2 m_3 {M_3}^3 {e_0}^2 \sin \iota \sin \iota_3 \sin 2\omega_0}{8 {J_3}^3
 \mu_0 {L_3}^3 M_0} \ ,  \nonumber  \\
& K^{\ast\ast}_{(0),0,4}=-\frac{15 \Delta m {\mu_3}^8 {L_0}^4 m_3 {M_3}^4 e_0 e_3 }{256 {J_3}^5 {\mu_0}^5 {L_3}^3 {M_0}^4} \big\{{J_0}^2 \big[\cos \omega_0 \cos \omega_3 \big(70 \sin^2 \iota \cos 2\omega_0+25 \cos 2\iota-13\big) \nonumber \\
& \hspace*{3cm} +\cos \iota \sin \omega_0 \sin \omega_3 \big(70 \sin^2 \iota \cos 2\omega_0+5 \cos 2\iota+7\big)\big]
 \nonumber \\
&\hspace*{2cm} -7 {L_0}^2 \big(10 \sin^2 \iota \cos 2\omega_0+5 \cos 2\iota-1\big) \big(\cos \iota \sin \omega_0 \sin \omega_3+\cos \omega_0 \cos \omega_3\big)\big\} \ ,
  \nonumber \\
&  \frac{\partial K^{\ast\ast}_{(0),0,4}}{\partial \delta \Omega}=\frac{15 \Delta m {\mu_3}^8 {L_0}^4 m_3 {M_3}^4 e_0 e_3 }{1024 {J_3}^5 {\mu_0}^5 {L_3}^3 {M_0}^4} \big\{\big(3 {J_0}^2-7 {L_0}^2\big) \big[\big(6 \cos \iota_0-5 \cos \big(\iota_0+2 \iota_3\big) \nonumber \\
& \hspace*{3cm}
  +15 \cos \big(3 \iota_0+2 \iota_3\big)\big) \sin \omega_0 \cos \omega_3+\big(5 \big(\cos \big(2 \iota_0+\iota_3\big)-3 \cos \big(2 \iota_0+3 \iota_3\big)\big) \nonumber \\
& \hspace*{3cm}  -6 \cos \iota_3\big) \sin \omega_3 \cos \omega_0\big] \nonumber \\
& \hspace*{2cm}
-35 a_0 {\mu_0}^2 M_0 {e_0}^2 \big[-\big(\big(-6 \cos \iota_0+5 \cos \big(\iota_0+2 \iota_3\big)+\cos \big(3 \iota_0+2 \iota_3\big)\big) \sin 3\omega_0 \cos \omega_3\big)
\nonumber \\
& \hspace*{3cm}
 -\big(2 \cos \iota_3+\cos \big(2 \iota_0+\iota_3\big)-3 \cos \big(2 \iota_0+3 \iota_3\big)\big) \sin \omega_3 \cos 3\omega_0\big]\big\}\ ,  \nonumber \\
& K^{\ast\ast}_{(1),0,9/2}=\frac{3 J_0 {\mu_3}^9 {L_0}^4 {m_3}^2 {M_3}^4 }{256 {J_3}^6 {\mu_0}^6 {L_3}^5 {M_0}^4 \big(J_3+L_3\big)} \big\{\big(J_3-L_3\big) \big(10 J_3 L_3+3 {J_3}^2+5 {L_3}^2\big) \cos 2\omega_3 \big[5 a_0 {\mu_0}^2 M_0 {e_0}^2 \nonumber \\
& \hspace*{3cm}
\times \big(5 \cos \iota+3 \cos 3\iota\big) \cos 2\omega_0+4 \sin^2 \iota \cos \iota \big(15 {L_0}^2-17 {J_0}^2\big)\big]
\nonumber \\
& \hspace*{2cm}
 +40 a_0 {\mu_0}^2 M_0 {e_0}^2 \cos 2\iota \big(J_3-L_3\big) \big(10 J_3 L_3+3 {J_3}^2+5 {L_3}^2\big) \sin 2\omega_0 \sin 2\omega_3 \nonumber \\
& \hspace*{3cm}
  -2 \cos \iota \big(J_3+L_3\big) \big(2 {J_3}^2-5 {L_3}^2\big) \big[{J_0}^2 \big(30 \sin^2 \iota \cos 2\omega_0+17 \cos 2\iota+83\big)\nonumber \\
& \hspace*{3cm}
  -5 {L_0}^2 \big(6 \sin^2 \iota \cos 2\omega_0+3
  \cos 2\iota+17\big)\big]\big\} \ ,
 \nonumber \\
& \frac{\partial K^{\ast\ast}_{(1),0,9/2}}{\partial \delta \Omega}=\frac{3 J_0 {\mu_3}^9 {L_0}^4 {m_3}^2 {M_3}^4 }{256 {J_3}^6 {\mu_0}^6 {L_3}^5 {M_0}^4 \big(J_3+L_3\big)} \big\{-60 a_0 {\mu_0}^2 M_0 {e_0}^2 \sin 2\iota \sin \iota_3 \big(J_3+L_3\big)
\big(2 {J_3}^2-5 {L_3}^2\big) \sin 2\omega_0
\nonumber \\
&\hspace*{2cm}
 -10 a_0 {\mu_0}^2 M_0 {e_0}^2 \big(5 \cos \big(2 \iota_0+\iota_3\big)+3 \cos \big(2 \iota_0+3 \iota_3\big)\big) \big(J_3-L_3\big) \big(10 J_3 L_3+3 {J_3}^2+5 {L_3}^2\big)
 \nonumber \\
 & \hspace*{3cm} \times \sin 2\omega_0 \cos 2\omega_3
 \nonumber \\
& \hspace*{2cm}
  +2 \big(J_3-L_3\big) \big(10 J_3 L_3+3 {J_3}^2+5 {L_3}^2\big) \sin 2\omega_3 \big[5 a_0 {\mu_0}^2 M_0 {e_0}^2 \big(5 \cos \big(\iota_0 +2 \iota_3\big)
   \nonumber \\
& \hspace*{3cm}
 +3 \cos \big(3 \iota_0
 +2 \iota_3\big)\big) \cos 2\omega_0
 +2 \sin 2\iota \sin  \iota_0 \big(15 {L_0}^2-17 {J_0}^2\big)\big]\big\} \ , \nonumber \\
& K^{\ast\ast}_{(0),1,1/2}=0 \ , \nonumber \\
& \frac{\partial K^{\ast\ast}_{(0),1,1/2}}{\partial \delta \Omega}=0 \ , \nonumber \\
&  K^{\ast\ast}_{(0),1,1}=\frac{\mu_0 {\mu_3}^2 M_3 }{4 a_0 {L_3}^2 M_0} \big(2 m_3 {M_0}^2+{\mu_3}^2 M_3\big)\ ,
  \nonumber \\
& \frac{\partial K^{\ast\ast}_{(0),1,1}}{\partial \delta \Omega}=0 \ , \nonumber \\
& K^{\ast\ast}_{(1),1,3/2}=0\ , \nonumber  \\
& \frac{\partial K^{\ast\ast}_{(1),1,3/2}}{\partial \delta \Omega}=0 \ , \nonumber \\
& K^{\ast\ast}_{(0),1,2}=0 \ , \nonumber  \\
& \frac{\partial K^{\ast\ast}_{(0),1,2}}{\partial \delta \Omega}=0 \ , \nonumber \\
& K^{\ast\ast}_{(0),1,5/2}=\frac{J_0 {\mu_3}^6 {M_3}^3 \cos \iota }{2 {J_3}^2 {L_3}^3 M_0} \big(3 m_3+4 M_0\big) \ ,
  \nonumber \\
& \frac{\partial K^{\ast\ast}_{(0),1,5/2}}{\partial \delta \Omega}=0 \ , \nonumber \\
& K^{\ast\ast}_{(1),1,5/2}=0 \ , \nonumber \\
& \frac{\partial K^{\ast\ast}_{(1),1,5/2}}{\partial \delta \Omega}=0 \ , \nonumber \\
& K^{\ast\ast}_{(0),1,3}=\frac{{\mu_3}^6 m_3 {M_3}^3 }{64 {J_3}^3 \mu_0 {L_3}^3 M_0} \big\{3 \big(M_0-\mu_0\big) \big[6 a_0 {\mu_0}^2 M_0 {e_0}^2 \sin^2 \iota \cos 2\omega_0+{L_0}^2 \big(3 \cos 2\iota+1\big)\big] \nonumber  \\
& \hspace*{2cm}  -{J_0}^2 \big(3 \cos 2\iota+1\big) \big(5 M_0-13 \mu_0\big)\big\}\ , \nonumber \\
& \frac{\partial K^{\ast\ast}_{(0),1,3}}{\partial \delta \Omega}=\frac{9 a_0 \mu_0 {\mu_3}^6 m_3 {M_3}^3 {e_0}^2 \sin \iota \sin \iota_3 \sin 2\omega_0}{16 {J_3}^3 {L_3}^3} \big(\mu_0-M_0\big) \ ,
\end{align}
where $a_0 = {L_0}^2/({\mu_0}^2 M_0)$,
  $e_0 = \sqrt{{L_0}^2-{J_0}^2}/L_0$,
  and $\iota_0 = \cos \left(H_0/J_0\right)$,
  and similarly for outer binaries.
Having mixed the orbital elements of semi-major axes, eccentricities, etc,
  with the canonical coordinates and momenta in our formula,
  the evolution equations can be derived from $K^{\ast\ast}$ using
\begin{align}
&    \frac{{\rm d} l_0}{{\rm d} t}  = \frac{\partial  K^{\ast\ast}}{\partial  L_0}  +
    \frac{2 L_0}{{\mu_0}^2 M_0}  \frac{\partial  K^{\ast\ast}}{\partial a_0} + \frac{{J_0}^2}{{L_0}^3 e_0} \frac{\partial  K^{\ast\ast}}{\partial  e_0} \ ,
     \nonumber \\
&    \frac{{\rm d} e_0}{{\rm d} t}  = \frac{J_0}{e_0 {L_0}^2} \frac{\partial  K^{\ast\ast}}{\partial  \omega_0} \ ,
 \nonumber \\
&    \frac{{\rm d} \iota_0}{{\rm d} t}  = - \frac{1}{J_0 \sin \iota_0} \left( \cos \iota_0 \frac{\partial  K^{\ast\ast}}{\partial  \omega_0}
     - \frac{\partial  K^{\ast\ast}}{\partial  \delta \Omega} \frac{\partial  \delta \Omega}{\partial  \Omega_0}
    \right) \ ,
      \nonumber  \\
&    \frac{{\rm d} \omega_0}{{\rm d} t}  = \frac{\partial  K^{\ast\ast}}{\partial  J_0} +
    \frac{\cot \iota_0 }{J_0} \frac{\partial  K^{\ast\ast}}{\partial  \iota}
    - \frac{J_0}{e_0 {L_0}^2} \frac{\partial  K^{\ast\ast}}{\partial  e_0} \ ,
    \nonumber \\
&    \frac{{\rm d} \Omega_0}{{\rm d} t}  = - \frac{1}{J_0 \sin \iota_0} \frac{\partial  K^{\ast\ast}}{\partial  \iota} \ ,
\end{align}
and similarly for the outer binary.
The evolution equations and detailed generating functions
  are available as {\it Mathematica} script at
  \url{https://github.com/kayejlli/Relativistic3Body}.

\twocolumngrid
\bibliography{apssamp}
\bibliographystyle{apsrev}

\end{document}